\pdfoutput=1

\documentclass[journal]{IEEEtran}
\usepackage{siunitx}
\usepackage{graphicx}
\usepackage{amsmath}
\usepackage{xcolor}

\usepackage[acronym,nomain]{glossaries}
\glsdisablehyper
\newcommand{\SetCapsType}{normalcaps}
\makeatletter
\providecommand{\SetCapsType}{smallcaps}

\long\def\@scTrue{smallcaps}
\long\def\@scFalse{normalcaps}
\newcommand{\acroSCaps}[1]{%
 \begingroup
  \ifx\SetCapsType\@scTrue 
    \textsc{#1}%
  \else
    \MakeUppercase{#1}%
  \fi
  \endgroup
}
\makeatother

\newcommand{\nAcronym}[4][]{%
	\newacronym[#1]{#2}{#3}{#4}
}

\makeatletter
\@ifpackageloaded{babel}{%
    \newcommand{\usuk}[2]{%
        \iflanguage{USenglish}{#1}{#2}%
    }%
}{%
    \newcommand{\usuk}[2]{%
        #1%
    }%
}%
\makeatother

\usepackage[shortcuts]{extdash}% prevent line break after hyphen before QAM
\newcommand{\qam}[1]{%, % at end of line result in no spaces before and after command
    \ifglsused{QAM}%
        {#1\=/\gls{QAM}}%
        {#1\=/ary \gls{QAM}%
    }%
}%
\let\QAM\qam

\newcommand{\pam}[1]{%, % at end of line result in no spaces before and after command
    \ifglsused{PAM}%
        {\gls{PAM}\=/#1}%
        {#1\=/ary \gls{PAM}%
    }%
}%
\let\PAM\pam

\nAcronym{2A8PSK}{\acroSCaps{2a8psk}}{2-ary amplitude 8-ary phaseshift keying}

\nAcronym{3CCMCF}{\acroSCaps{3cc-mcf}}{3-core coupled-core multi-core fiber}

\nAcronym{4D}{\acroSCaps{4d}}{four-dimensional}
\nAcronym{4D64PRS}{\acroSCaps{4d-64prs}}{\usuk{four-dimensional 64-ary polarization-ring-switching}{four-dimensional 64-ary polarisation-ring-switching}}

\nAcronym{5B4D2A8PSK}{\acroSCaps{5b4d-2a8psk}}{5-bit four-dimensional two-amplitude 8-ary phase-shift keying}

\nAcronym{6B4D2A8PSK}{\acroSCaps{6b4d-2a8psk}}{6-bit four-dimensional two-amplitude 8-ary phase-shift keying}

\nAcronym{8D}{\acroSCaps{8d}}{eight-dimensional}\nAcronym{8D2048PRS}{\acroSCaps{8d-2048prs}}{eight-dimensional 2048-ary polarization-ring-switching}
\nAcronym{8D2048PRST1}{\acroSCaps{8d-2048prs-t1}}{eight-dimensional 2048-ary polarization-ring-switching type 1}
\nAcronym{8D2048PRST2}{\acroSCaps{8d-2048prs-t2}}{eight-dimensional 2048-ary polarization-ring-switching type 2}

\nAcronym{ABC}{\acroSCaps{abc}}{automatic bias controller}
\nAcronym{ADC}{\acroSCaps{adc}}{analog-to-digital converter}
\nAcronym{AIR}{\acroSCaps{air}}{achievable information rate}
\nAcronym{AOM}{\acroSCaps{aom}}{acoustic optical modulator}
\nAcronym{API}{\acroSCaps{api}}{application programming interface}
\nAcronym{AR}{\acroSCaps{ar}}{achievable rate}
\nAcronym{ASE}{\acroSCaps{ase}}{amplified spontaneous emission}
\nAcronym{ASK}{\acroSCaps{ask}}{amplitude-shift keying}
\nAcronym{ASIC}{\acroSCaps{asic}}{application-specific integrated circuit}
\nAcronym{ARRWG}{\acroSCaps{a}rr\acroSCaps{wg}}{arrayed-waveguide grating}
\nAcronym{AWG}{\acroSCaps{awg}}{arbitrary-waveform generator}
\nAcronym{AWGN}{\acroSCaps{awgn}}{additive white Gaussian noise}

\nAcronym{BCH}{\acroSCaps{bch}}{Bose-Chaudhuri-Hocquenghem}
\nAcronym{BER}{\acroSCaps{ber}}{bit error rate}
\nAcronym{BICM}{\acroSCaps{bicm}}{bit-interleaved coded modulation}
\nAcronym{BMD}{\acroSCaps{bmd}}{bit-metric decoding}
\nAcronym{BPD}{\acroSCaps{bpd}}{balanced photo-diode}
\nAcronym{BPF}{\acroSCaps{bpf}}{bandpass filter}
\nAcronym{BPS}{\acroSCaps{bps}}{blind phase search}
\nAcronym{BPSK}{\acroSCaps{bpsk}}{binary phase-shift keying}
\nAcronym{BRGC}{\acroSCaps{brgc}}{binary reflected Gray code}
\nAcronym{BTB}{\acroSCaps{btb}}{back-to-back}

\nAcronym{CAGR}{\acroSCaps{cagr}}{compound annual growth rate}
\nAcronym{CCDM}{\acroSCaps{ccdm}}{constant composition distribution matching}
\nAcronym{CCF}{\acroSCaps{ccf}}{\usuk{coupled-core fiber}{coupled-core fibre}}
\nAcronym{CD}{\acroSCaps{cd}}{chromatic dispersion}
\nAcronym{CIR}{\acroSCaps{cir}}{channel impulse response}
\nAcronym{CMA}{\acroSCaps{cma}}{constant modulus algorithm}
\nAcronym{CMUX}{\acroSCaps{cmux}}{core multiplexer}
\nAcronym{ChUT}{\acroSCaps{chut}}{channel under test}
\nAcronym{CUT}{\acroSCaps{cut}}{channel under test}
\nAcronym{CPE}{\acroSCaps{cpe}}{carrier phase estimation}
\nAcronym{CPU}{\acroSCaps{cpu}}{central processing unit}
\nAcronym{CSPR}{\acroSCaps{cspr}}{carrier-to-signal power ratio}
\nAcronym{CW}{\acroSCaps{cw}}{continuous wave}

\nAcronym{DA}{\acroSCaps{da}}{driver amplifier}
\nAcronym{DAC}{\acroSCaps{dac}}{digital-to-analog converter}
\nAcronym{DCF}{\acroSCaps{dcf}}{\usuk{dispersion compensated fiber}{dispersion compensated fibre}}
\nAcronym{DCI}{\acroSCaps{dci}}{data center interconnect}
\nAcronym{DDLMS}{\acroSCaps{dd-lms}}{decision-directed least mean square}
\nAcronym{DFB}{\acroSCaps{dfb}}{distributed feedback}
\nAcronym{DGD}{\acroSCaps{dgd}}{differential group delay}
\nAcronym{DH}{\acroSCaps{dh}}{digital holography}
\nAcronym{DM}{\acroSCaps{dm}}{distribution matcher}
\nAcronym{DMA}{\acroSCaps{dma}}{direct memory access}
\nAcronym{DMD}{\acroSCaps{dmd}}{differential mode delay}
\nAcronym{DMG}{\acroSCaps{dmg}}{differential modal gain}
\nAcronym{DMGD}{\acroSCaps{dmgd}}{differential mode group delay}
\nAcronym{DML}{\acroSCaps{dml}}{directly-modulated laser}
\nAcronym{DP}{\acroSCaps{dp}}{\usuk{dual-polarization}{dual-polarisation}}
\nAcronym{DPC}{\acroSCaps{dpc}}{digital pre-compensation}
\nAcronym{DPE}{\acroSCaps{dpe}}{digital pre-emphasis}
\nAcronym{DPIQ}{\acroSCaps{dp-iqm}}{\usuk{dual-polarization IQ-modulator}{dual-polarisation IQ-modulator}}
\nAcronym{DPLL}{\acroSCaps{dpll}}{digital phase-locked loop}
\nAcronym{DQPSK}{\acroSCaps{dqpsk}}{differential quaternary phase-shift-keying}
\nAcronym{DRA}{\acroSCaps{dra}}{distributed Raman amplifier}
\nAcronym{DRE}{\acroSCaps{dre}}{digital resolution enhancer}
\nAcronym{DSF}{\acroSCaps{dsf}}{\usuk{dispersion-shifted fiber}{dispersion-shifted fibre}}
\nAcronym{DSO}{\acroSCaps{dso}}{digital sampling oscilloscope}
\nAcronym{DSP}{\acroSCaps{dsp}}{digital signal processing}
\nAcronym{DUT}{\acroSCaps{dut}}{device-under-test}
\nAcronym{DWDM}{\acroSCaps{dwdm}}{dense wavelength-division multiplexing}

\nAcronym{EAM}{\acroSCaps{eam}}{electro-absorption modulator}
\nAcronym{ECL}{\acroSCaps{ecl}}{external cavity laser}
\nAcronym{ED}{\acroSCaps{ed}}{Eucledian distance}
\nAcronym{EDFA}{\acroSCaps{edfa}}{\usuk{erbium-doped fiber amplifier}{erbium-doped fibre amplifier}}
\nAcronym{ENOB}{\acroSCaps{enob}}{effective number of bits}
\nAcronym{ESS}{\acroSCaps{ess}}{enumerative sphere shaping}

\nAcronym{FD}{\acroSCaps{fd}}{frequency domain}
\nAcronym{FDE}{\acroSCaps{fde}}{\usuk{frequency domain equalizer}{frequency domain equaliser}}
\nAcronym{FEC}{\acroSCaps{fec}}{forward error correction}
\nAcronym{FFT}{\acroSCaps{fft}}{fast Fourier transform}
\nAcronym{FIR}{\acroSCaps{fir}}{finite impulse response}
\nAcronym{FMF}{\acroSCaps{fmf}}{\usuk{few-mode fiber}{few-mode fibre}}
\nAcronym[plural=FM-MCF, firstplural=\usuk{few-mode multi-core fibers}{few-mode multi-core fibres}]{FM-MCF}{\acroSCaps{fm-mcf}}{\usuk{few-mode multi-core fiber}{few-mode multi-core fibre}}
\nAcronym{FPGA}{\acroSCaps{fpga}}{field-programmable gate array}
\nAcronym{FWM}{\acroSCaps{fwm}}{four-wave mixing}

\nAcronym{GD}{\acroSCaps{gd}}{group delay}
\nAcronym{GI}{\acroSCaps{gi}}{graded-index}
\nAcronym{GFF}{\acroSCaps{gff}}{gain flattening filter}
\nAcronym{GIMMF}{\acroSCaps{gi-mmf}}{\usuk{graded-index multi-mode fiber}{graded-index multi-mode fibre}}
\nAcronym{GMI}{\acroSCaps{gmi}}{\usuk{generalized mutual information}{generalised mutual information}}
\nAcronym{GNSE}{\acroSCaps{gnse}}{generalized nonlinear Schr\"{o}dinger equation}
\nAcronym{GPU}{\acroSCaps{gpu}}{graphics processing unit}
\nAcronym{GS}{\acroSCaps{gs}}{geometric shaping}
\nAcronym{GV}{\acroSCaps{gv}}{group-velocity}
\nAcronym{GVD}{\acroSCaps{gvd}}{group-velocity dispersion}
\nAcronym{GPIO}{\acroSCaps{gpio}}{general purpose input output}

\nAcronym{HDFEC}{\acroSCaps{hd-fec}}{hard decision forward error correction}

\nAcronym[plural=IL, firstplural=insertion losses (\textsc{il})]{IL}{\acroSCaps{il}}{insertion loss}
\nAcronym{IFFT}{\acroSCaps{ifft}}{inverse fast Fourier transform}
\nAcronym{IMDD}{\acroSCaps{im/dd}}{intensity-modulation direct-detection}
\nAcronym{IQM}{\acroSCaps{iqm}}{in-phase and quadrature modulator}
\nAcronym{ISI}{\acroSCaps{isi}}{intersymbol interference}

\nAcronym{KK}{\acroSCaps{kk}}{Kramers-Kronig}

\nAcronym{LCOS}{\acroSCaps{LCoS}}{liquid crystal on silicon}
\nAcronym{LDPC}{\acroSCaps{ldpc}}{low-density parity-check}
\nAcronym{LEAF}{\acroSCaps{leaf}}{\usuk{large effective area fiber}{large effective area fibre}}
\nAcronym{LFSR}{\acroSCaps{lfsr}}{linear-feedback shift register}
\nAcronym{LMS}{\acroSCaps{lms}}{least means square}
\nAcronym{LLR}{\acroSCaps{llr}}{log-likelihood ratio}
\nAcronym{LO}{\acroSCaps{lo}}{local oscillator}
\nAcronym{LP}{\acroSCaps{lp}}{\usuk{linearly polarized}{linearly polarised}}
\nAcronym{LSPS}{\acroSCaps{lsps}}{\usuk{loop-synchronized polarization scrambler}{loop-synchronised polarisation scrambler}}
\nAcronym{LUT}{\acroSCaps{lut}}{lookup table}

\nAcronym{MB}{\acroSCaps{mb}}{Maxwell-Bolzmann}
\nAcronym{MCF}{\acroSCaps{mcf}}{\usuk{multi-core fiber}{multi-core fibre}}
\nAcronym{MDG}{\acroSCaps{mdg}}{mode dependent gain}
\nAcronym[plural=MDL, firstplural=mode-dependent losses (\textsc{mdl})]{MDL}{\acroSCaps{mdl}}{mode-dependent loss}
\nAcronym{MDM}{\acroSCaps{mdm}}{mode division multiplexing}
\nAcronym{MEMS}{\acroSCaps{mems}}{micro-electro-mechanical systems}
\nAcronym{MF}{\acroSCaps{mf}}{matched filter}
\nAcronym{MI}{\acroSCaps{mi}}{mutual information}
\nAcronym{MIMO}{\acroSCaps{mimo}}{multiple-input multiple-output}
\nAcronym{ML}{\acroSCaps{ml}}{machine learning}
\nAcronym{MMA}{\acroSCaps{mma}}{multi-modulus algorithm}
\nAcronym{MMF}{\acroSCaps{mmf}}{\usuk{multi-mode fiber}{multi-mode fibre}}
\nAcronym{MMSE}{\acroSCaps{mmse}}{minimum mean squared error}
\nAcronym{MPLC}{\acroSCaps{mplc}}{multi-plane light converter}
\nAcronym{MSE}{\acroSCaps{mse}}{mean squared error}
\nAcronym{MUX}{\acroSCaps{mux}}{multiplexer}
\nAcronym{MZM}{\acroSCaps{mzm}}{Mach-Zehnder modulator}
\nAcronym{MZI}{\acroSCaps{mzi}}{Mach-Zehnder interferometer}

\nAcronym{NA}{\acroSCaps{na}}{numerical aperture}
\nAcronym{NF}{\acroSCaps{nf}}{noise figure}
\nAcronym{NGMI}{\acroSCaps{ngmi}}{\usuk{normalized generalized mutual information}{normalised generalised mutual information}}
\nAcronym{NLSE}{\acroSCaps{nlse}}{nonlinear Schr\"{o}ding equation}
\nAcronym{NIR}{\acroSCaps{nir}}{near-infrared}

\nAcronym{OBTB}{\acroSCaps{obtb}}{optical back-to-back}
\nAcronym{ODE}{\acroSCaps{ode}}{ordinary differential equation}
\nAcronym{ODL}{\acroSCaps{odl}}{optical delay line}
\nAcronym{OFC}{\acroSCaps{ofc}}{Optical Fiber Communications Conference}
\nAcronym{OFDR}{\acroSCaps{ofdr}}{optical frequency-domain reflectometer}
\nAcronym{OFDM}{\acroSCaps{ofdm}}{orthogonal frequency division multiplexing}
\nAcronym{OMFT}{\acroSCaps{omft}}{optical multi-format transmitter}
\nAcronym{OOK}{\acroSCaps{ook}}{on-off keying}
\nAcronym{OPLL}{\acroSCaps{opll}}{optical phase-locked loop}
\nAcronym{OSA}{\acroSCaps{osa}}{\usuk{optical spectrum analyzer}{optical spectrum analyser}}
\nAcronym{OSNR}{\acroSCaps{osnr}}{optical signal-to-noise ratio}
\nAcronym{OTDR}{\acroSCaps{otdr}}{optical time-domain reflectometer}
\nAcronym{OTF}{\acroSCaps{otf}}{optical tunable filter}
\nAcronym{OVNA}{\acroSCaps{ovna}}{\usuk{optical vector network analyzer}{optical vector network analyser}}

\nAcronym{PAM}{\acroSCaps{pam}}{pulse-amplitude modulation}
\nAcronym{PAS}{\acroSCaps{pas}}{probabilistic amplitude shaping}
\nAcronym{PAPR}{\acroSCaps{papr}}{peak-to-avarage power ratio}
\nAcronym{PBC}{\acroSCaps{pbc}}{\usuk{polarization beam combiner}{polarisation beam combiner}}
\nAcronym{PBS}{\acroSCaps{pbs}}{polarization beam splitter}
\nAcronym{PCVD}{\acroSCaps{pcvd}}{plasma chemical vapor depostion}
\nAcronym{PD}{\acroSCaps{pd}}{photodiode}
\nAcronym{PDL}{\acroSCaps{pdl}}{polarization-dependent loss}
\nAcronym{PDM}{\acroSCaps{pdm}}{\usuk{polarization-division multiplexing}{polarisation-division multiplexing}}
\nAcronym{PL}{\acroSCaps{pl}}{photonic lantern}
\nAcronym{PMBPSK}{\acroSCaps{pm-bpsk}}{polarization-multiplexed binary phase-shift-keying}
\nAcronym{PMQPSK}{\acroSCaps{pm-qpsk}}{polarization-multiplexed quaternary phase-shift-keying}
\nAcronym{PM8QAM}{\acroSCaps{pm-8qam}}{polarization-multiplexed 8-ary quadrature amplitude modulation}
\nAcronym{PMD}{\acroSCaps{pmd}}{polarization mode dispersion}
\nAcronym{PNOB}{\acroSCaps{pnob}}{physical number of bits}
\nAcronym{PON}{\acroSCaps{pon}}{passive-optical network}
\nAcronym{PRBS}{\acroSCaps{prbs}}{pseudorandom bit sequence}
\nAcronym{PS}{\acroSCaps{ps}}{probabilistic shaping}
\nAcronym{PSK}{\acroSCaps{psk}}{phase-shift keying}
\nAcronym{PSP}{\acroSCaps{psp}}{\usuk{principal states of polarization}{principal states of polarisation}}

\nAcronym{QAM}{\acroSCaps{qam}}{quadrature amplitude modulation}
\nAcronym{QPSK}{\acroSCaps{qpsk}}{quaternary phase-shift-keying}

\nAcronym{RAM}{\acroSCaps{ram}}{random-access memory}
\nAcronym{RF}{\acroSCaps{rf}}{radio frequency}
\nAcronym{RRC}{\acroSCaps{rrc}}{root-raised-cosine}
\nAcronym{ROADM}{\acroSCaps{roadm}}{reconfigurable optical add-drop multiplexer}
\nAcronym{ROI}{\acroSCaps{roi}}{region of interest}

\nAcronym{SA}{\acroSCaps{sa}}{simulated annealing}
\nAcronym{SamPerSym}{\acroSCaps{sps}}{samples per symbol}
\nAcronym{SBS}{\acroSCaps{sbs}}{stimulated Brillouin scattering}
\nAcronym{SDFEC}{\acroSCaps{sd-fec}}{soft decision forward error correction}
\nAcronym{SDM}{\acroSCaps{sdm}}{space-division multiplexing}
\nAcronym{SE}{\acroSCaps{se}}{spectral efficiency}
\nAcronym{SER}{\acroSCaps{ser}}{symbol error rate}
\nAcronym{SI}{\acroSCaps{si}}{step index}
\nAcronym{SLM}{\acroSCaps{slm}}{spatial light modulator}
\nAcronym{SMF}{\acroSCaps{smf}}{\usuk{single-mode fiber}{single-mode fibre}}
\nAcronym{SNR}{\acroSCaps{snr}}{signal-to-noise ratio}
\nAcronym{SOA}{\acroSCaps{soa}}{semiconductor optical amplifier}
\nAcronym[firstplural=\usuk{states of polarization (\acroSCaps{sop})}{states of polarisation (\acroSCaps{sop})}]{SOP}{\acroSCaps{sop}}{\usuk{state of polarization}{state of polarisation}}
\nAcronym{SPM}{\acroSCaps{spm}}{self-phase modulation}
\nAcronym{SPS}{\acroSCaps{sps}}{\usuk{synchronized polarization scrambler}{synchronised polarisation scrambler}}
\nAcronym{SRS}{\acroSCaps{srs}}{stimulated Raman scattering}
\nAcronym{SSBI}{\acroSCaps{ssbi}}{signal-signal beat interference}
\nAcronym{SSFM}{\acroSCaps{ssfm}}{split-step Fourier method}
\nAcronym{SSMF}{\acroSCaps{ssmf}}{\usuk{standard single-mode fiber}{standard single-mode fibre}}
\nAcronym{STL}{\acroSCaps{stl}}{swept tunable laser}
\nAcronym{SVD}{\acroSCaps{svd}}{singular value decomposition}
\nAcronym{SWI}{\acroSCaps{swi}}{swept wavelength interferometry}

\nAcronym{TD}{\acroSCaps{td}}{time domain}
\nAcronym{TDE}{\acroSCaps{tde}}{\usuk{time domain equalizer}{time domain equaliser}}
\nAcronym{TDMSDM}{\acroSCaps{tdm-sdm}}{time-domain multiplexed space-division multiplexing}
\nAcronym{TE}{\acroSCaps{te}}{transverse electric}
\nAcronym{TIA}{\acroSCaps{tia}}{trans-impedance amplifier}
\nAcronym{TM}{\acroSCaps{TM}}{transverse magnetic}
\nAcronym{TH4D}{\acroSCaps{th-4d}}{time domain hybrid four-dimensional}
\nAcronym{TH4D2A8PSK}{\acroSCaps{th-4d-2a8psk}}{time domain hybrid four-dimensional two-amplitude eight-phase shift keying}

\nAcronym{VCSEL}{\acroSCaps{vcsel}}{vertical-cavity surface emitting laser}
\nAcronym{VOA}{\acroSCaps{voa}}{variable optical attenuator}

\nAcronym{WDM}{\acroSCaps{wdm}}{wavelength-division multiplexing}
\nAcronym{WGA}{\acroSCaps{wga}}{weakly guiding approximation}
\nAcronym{WGN}{\acroSCaps{wgn}}{white Gaussian noise}
\nAcronym{WSS}{\acroSCaps{wss}}{wavelength selective switch}

\nAcronym{XPM}{\acroSCaps{xpm}}{cross-phase modulation}
\nAcronym{XT}{\acroSCaps{xt}}{cross-talk}

\nAcronym{3DWG}{\acroSCaps{3dwg}}{3D-waveguide}

\usepackage[style=ieee, maxnames=3, defernumbers=true]{biblatex}
\bibliography{ref} 

\DeclareBibliographyCategory{ignore}

\usepackage{lipsum}

\usepackage[capitalise]{cleveref}

\usepackage{caption}
\usepackage{subcaption}
\ifCLASSINFOpdf
\else
\fi

\makeatletter
\let\blx@rerun@biber\relax
\makeatother

\begin{document}
% \title{Real-time Software-defined GPU-based optical receiver}
\title{Field Trial of a Flexible Real-time Software-defined GPU-based Optical Receiver}

\author{Sjoerd~van~der~Heide,~\IEEEmembership{Student Member,~IEEE,}
        Ruben~S.~Luis,~\IEEEmembership{Senior Member,~IEEE,}
        Benjamin~J.~Puttnam,~\IEEEmembership{Member,~IEEE,}
        Georg~Rademacher,~\IEEEmembership{Senior Member,~IEEE,}
        Ton~Koonen,~\IEEEmembership{Fellow,~IEEE,}
        Satoshi~Shinada,~\IEEEmembership{Member,~IEEE,}
        Yohinari~Awaji,~\IEEEmembership{Member,~IEEE,}
        Hideaki~Furukawa,~\IEEEmembership{Member,~IEEE,}
        Chigo~Okonkwo,~\IEEEmembership{Senior Member,~IEEE}% <-this % stops a space
% \thanks{M. Shell was with the Department
% of Electrical and Computer Engineering, Georgia Institute of Technology, Atlanta,
% GA, 30332 USA e-mail: (see http://www.michaelshell.org/contact.html).}% <-this % stops a space
% \thanks{J. Doe and J. Doe are with Anonymous University.}% <-this % stops a space
% \thanks{Manuscript received April 19, 2005; revised August 26, 2015.}}

\thanks{Sjoerd~van~der~Heide, Ton~Koonen, and Chigo~Okonkwo are with the High Capacity Optical Transmission Laboratory, Electro-Optical Communications Group, Eindhoven University of Technology, PO Box 513, 5600 MB, Eindhoven, The Netherlands. (e-mail: s.p.v.d.heide@tue.nl, c.m.okonkwo@tue.nl, a.m.j.koonen@tue.nl). Ruben~S.~Luis, Benjamin~J.~Puttnam, Georg~Rademacher, Satoshi~Shinada, Yohinari~Awaji, and Hideaki~Furukawa are with the National Institute of Information and Communications Technology, Photonic Network System Laboratory, 4-2-1, Nukui-Kitamachi, Koganei, Tokyo, 184-8795,  Japan (e-mail: rluis@nict.go.jp)}%
\thanks{Manuscript received 2020-11-23; revised 2020-12-28; accepted 2021-01-05. Date of publication xxxx. This work was partly supported by the Dutch NWO Gravitation Program on Research Center for Integrated Nanophotonics under Grant GA~024.002.033.}
}


% \markboth{Journal of \LaTeX\ Class Files,~Vol.~14, No.~8, August~2015}%
% {Shell \MakeLowercase{\textit{et al.}}: Bare Demo of IEEEtran.cls for IEEE Journals}

\markboth{Journal of Lightwave Technology, Vol. X, No. X, Month Day, 2021. Copyright IEEE Journal of Lightwave Technology}{TBA}%



\maketitle

\begin{abstract}
We introduce a flexible, software-defined real-time multi-modulation format receiver implemented on an off-the-shelf general-purpose \gls{GPU}. The flexible receiver is able to process 2~GBaud 2-, 4-, 8-, and 16-ary \gls{PAM} signals as well as 1~GBaud 4-, 16- and 64-ary \gls{QAM} signals, with the latter detected using a \gls{KK} coherent receiver. Experimental performance evaluation is shown for back-to-back. In addition, by using the JGN high speed R\&D network testbed, performance is evaluated after transmission over 91~km field-deployed optical fiber and \glspl{ROADM}.
\end{abstract}


% \fi
\IEEEpeerreviewmaketitle

\section{Introduction}
\label{sec:introduction}

With the continual increase in demand for data-traffic at lower cost-per-bit, there is an increased interest in low-cost optical transceivers for data-center interconnects. Multi-vendor standards, e.g. \cite{OIF400ZR}, are key to the development and roll-out of these systems. Software-defined transceivers have supported and enhanced the widespread development of 5G and other wireless communications standards \cite{8949466}. These systems perform \gls{DSP} wholly \cite{GNURadio} or partially \cite{kazaz2017hardware} using off-the-shelf general purpose hardware, leading to high flexibility, combined with low development effort and rapid turnaround. Therefore, software-defined transceivers are expected to play an increasing role in the rapid development, validation, and test of optical communication standards.

Whilst commonplace for wireless systems, the development of software-defined transceivers for optical communications has been restricted by energy and computing power limitations. Recently, exploiting \glspl{FPGA} for real-time \gls{DSP} for optical communications has been investigated \cite{Randel_FPGA_2015, Beppu_FPGA_2020, Beppu_2_FPGA_2020}. GPU-based systems for optical communications are restricted by energy and computing power to prototype and test. With 45\% \cite{winzer_scaling_2017} year-on-year growth of computing power and 25\% increase \cite{sun_summarizing_2019} in energy efficiency (FLOPS per Watt), general-purpose \glspl{GPU} have the potential to meet demanding processing requirements. Note that, \gls{GPU} power efficiency showed a 3-fold improvement over equivalent \gls{FPGA} for simple highly-parallelized operations \cite{qasaimeh_comparing_2019}. These exponential increases may facilitate GPU use beyond prototyping. Compared to \glspl{GPU}, \glspl{FPGA} require longer development times and more stringent resource management to achieve the specific functions required for \gls{DSP}. 

Recently, the use of general-purpose \glspl{GPU} has been demonstrated for specific functions such as \gls{FEC} decoding \cite{Li_realtime_LDPC, suzuki_fec} and physical-layer functions for optical communications \cite{suzuki_phy_2018, suzuki_phy_2019, suzuki_phy_2020}. Additionally, real-time \gls{DSP} for optical \gls{DQPSK} has been implemented on a \gls{GPU} \cite{kim_OFC_2018, kim_JOCN_2019, suzuki_real-time_2020}. In these papers, massive parallel processing capabilities of \glspl{GPU} were exploited for processing single-polarization \SI{5}{Gbit/s} \gls{DQPSK} signals, correcting for \gls{ISI} using a \gls{FIR} filter. This approach greatly increases flexibility of optical transceivers. However, there remains the potential to further improve on this concept, since single-polarization coherent systems require real-time polarization control and differential phase-shift keyed modulation does not provide high spectral efficiency.

In this work, we implement a flexible,  software-defined real-time multi-modulation format receiver. A full real-time \gls{DSP} chain is implemented on a commercial, off-the-shelf general-purpose \gls{GPU} and validated experimentally. The receiver \gls{DSP} uses massive parallelization to receive \pam{2, -4, -8, and -16} signals at \SI{2}{GBaud} as well as \qam{4-, 16-, and 64} signals at \SI{1}{GBaud}, with the latter detected using a \gls{KK} coherent receiver\cite{mecozzi_kramers_2016}. All measurements employ identical transmitter and receiver hardware without polarization control. The \gls{GPU} software is able to switch between modulation formats. To the authors' knowledge, this is the first demonstration of a multi-modulation format software-defined GPU-based receiver and the first real-time demonstration of coherent \gls{KK} detection.

%%%%%%%%%%%%%%%%%%%%%%%%%%% Figure for the next section
\begin{figure*}
\includegraphics[width=\linewidth]{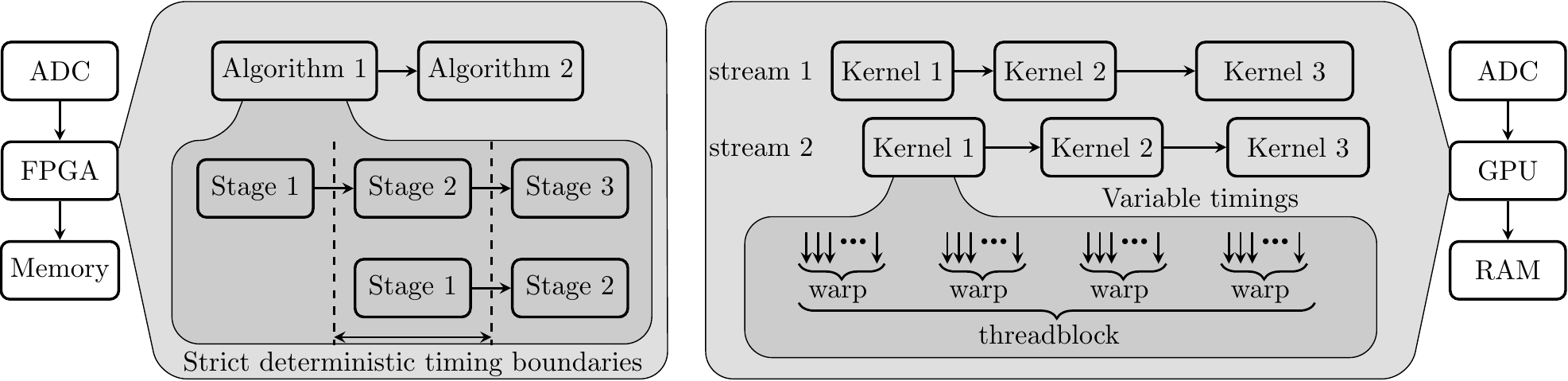}
\caption{Comparison between parallel processing on an FPGA (left) and a GPU (right). GPU processing algorithms are implemented in kernels which are executed in order in streams. Many threads in parallel perform the operations defined in the kernel. Threads are grouped in warps and threadblocks. Execution times on FPGAs are deterministic and strictly controlled, whilst timings on GPUs are non-deterministic.}
\label{fig:fpgagpu}
\end{figure*}

Furthermore, we validate the performance in a \SI{91}{km} optical fiber link over a field-deployed metropolitan network. The fiber ring is part of the Japan Gigabit Network (JGN) high speed R\&D network testbed \cite{JGN} consisting of 3 commercial \glspl{ROADM} in 2 separate Tokyo locations. These results demonstrate the potential of software-defined receivers for low-cost optical links, exploiting the exponentially growing computing power of \glspl{GPU}.

This paper is an extension to the work presented at the European Conference on Optical Communications (ECOC) 2020 \cite{ECOC_realtimeGPU}. Additional results and a detailed description of the structure of the real-time receiver architecture and the algorithms implemented on the \gls{GPU} are presented. Clock-recovery for \gls{IMDD} \pam{N} signals is shown to tolerate rapid changes in clock-frequency offset and static clock-frequency offsets of up to \SI{30.5}{ppm}. Using a noise-loading optical setup, \SI{2}{GBaud} \pam{2, 4, 8} signals are shown to reach the \SI{8.4}{dB} \gls{OSNR} Q-factor threshold for 6.7\% overhead \gls{HDFEC} \cite{agrell_information-theoretic_2018} at \SI{5.6}{dB}, \SI{14.0}{dB}, and \SI{22.2}{dB}, respectively. After transmission through the field trial network, an \gls{OSNR} penalty of \SI{0.4}{dB} and \SI{1.5}{dB} is observed for \pam{2} and \pam{4}, respectively. For \pam{8}, a 20\% overhead \gls{HDFEC} was necessary since it cannot reach the 6.7\% threshold. \pam{16} can be decoded in real time both in back-to-back and after transmission using the \gls{GPU} \gls{DSP}, but signal quality is not sufficient to reach either \gls{HDFEC} threshold. For \SI{1}{GBaud} \gls{KK} \qam{N} signals, \gls{CSPR} optimization was performed and a \gls{CSPR} of \SI{6}{dB} was chosen for \qam{4} and \SI{11}{dB} for \qam{16- and 64}. \qam{4- and 16} signals reach the 6.7\% overhead \gls{HDFEC} threshold at \SIlist{5.5;5.9}{dB} \gls{OSNR}, whilst \qam{16} requires an \gls{OSNR} of \SIlist{17.6; 19.1}{dB} for back-to-back and transmission, respectively. \qam{64} signals were processed in real time, but performance was not sufficient to reach either \gls{HDFEC} threshold. Six second continuous real-time transmission of all modulation formats show stable short-term average Q-factors despite the varying environment of installed fiber.

% Small change to win a line
This paper is structured as follows: \cref{sec:rxarch} introduces \gls{GPU} processing and the general structure of the real-time \gls{GPU} receiver architecture. \cref{sec:imdd} describes the \gls{DSP} algorithms employed for \gls{IMDD} \pam{N} signals in detail and with performance evaluation the in a back-to-back scenario. \cref{sec:kk} discusses the implementation and back-to-back evaluation for \gls{KK} \qam{N} signals. \cref{sec:results} discusses the evaluation of the real-time receiver evaluated using the experimental field trial network. Finally, \cref{sec:conclusion} concludes this paper.

\section{Real-time GPU Receiver Architecture}
\label{sec:rxarch}

\begin{figure*}
\includegraphics[width=\linewidth]{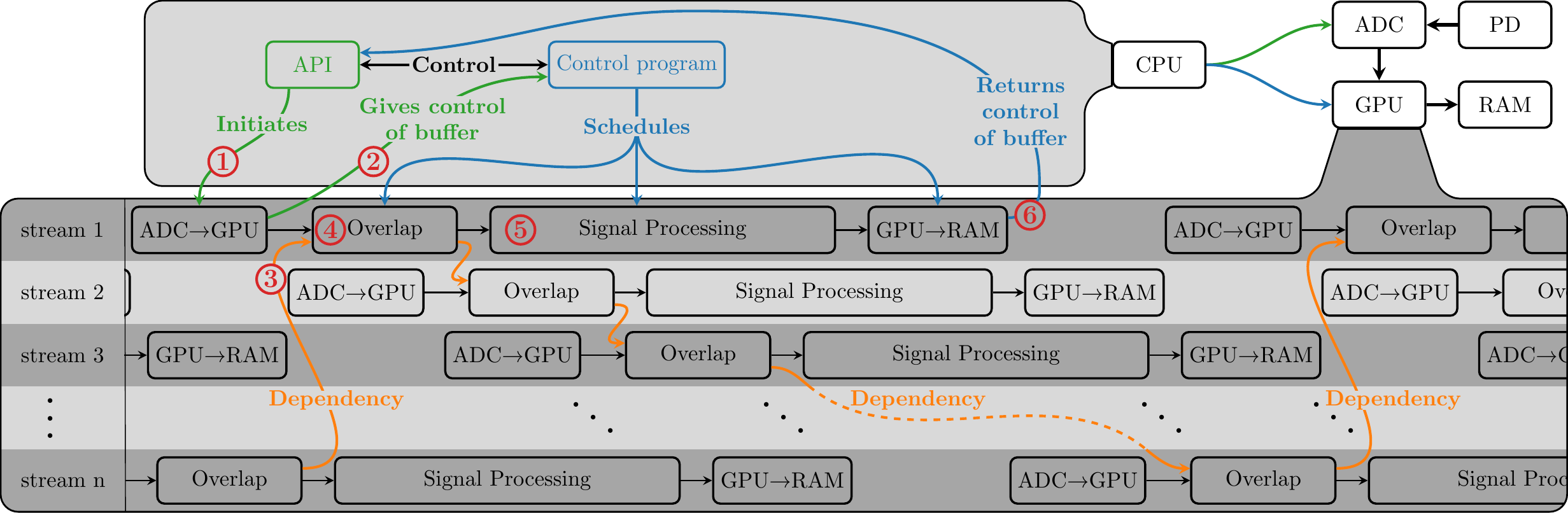}
\caption{Real-time \gls{GPU} Receiver Architecture. The output of the \gls{PD} is digitized by the \gls{ADC}. Samples are processed per buffer in a block-wise fashion in parallel streams on the \gls{GPU} controlled by the \gls{CPU}. Detailed descriptions of the \textit{Signal Processing} block can be found in \cref{sec:imdd,sec:kk}.}
\label{fig:rxarch}
\end{figure*}

\begin{figure}
\includegraphics[width=\linewidth]{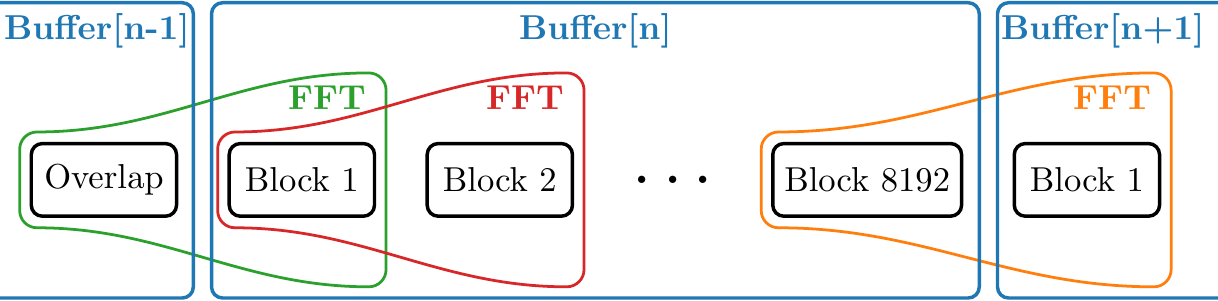}
\caption{Relation between buffers and blocks for block-wise processing. Each stream processes one buffer at a time, each consisting of 8192 blocks. Overlap-save Fourier transforms are used to ensure data integrity.}
\label{fig:dataarch}
\end{figure}

\subsection{Comparison between FPGA and GPU processing}
\cref{fig:fpgagpu} shows the similarities and differences between \gls{FPGA} and \gls{GPU} parallel processing architectures. Data are digitized by an \gls{ADC}, copied to the processing device, and after processing the results are stored in memory. Timings between parallel stages of processing in \glspl{FPGA} processing are deterministic and strictly controlled. The \gls{FPGA} operates at a certain clock rate and every stage of processing should fit within the timing parameters imposed by this central clock. Also, each stage of processing is assigned a fixed portion of physical computing hardware. In contrast, execution times on the \gls{GPU} are not deterministic. Computing hardware is shared for all kernels and a central scheduler assigns computing resources to kernels running in parallel. 

\subsection{GPU processing terminology}
Kernels are highly parallel routines that act upon data in the \gls{GPU} memory. The \textit{\gls{GPU} code} of the kernel is performed by \textit{threads} running in parallel. A thread is executed on a \gls{GPU} core and efficient implementations can use millions of threads.  A group of 32 threads is called a warp and is guaranteed to execute simultaneously, which allows for very efficient data exchange between these threads through \textit{warp-level shuffles}, used in this work for certain reduction kernels. A group of warps, called a \textit{threadblock}, is executed on the same \textit{streaming multiprocessor}, which is a group of \gls{GPU} cores. Threads in a threadblock share physical computing hardware and memory, leading to caching benefits. Multiple threadblocks are not necessarily performed in parallel. This depends on the scheduling by the \gls{GPU} driver.

Dependencies in the signal processing chain need to be handled appropriately. Kernels in the same processing \textit{stream} are performed in order. Therefore, splitting an algorithm into separate kernels in the same stream can address the dependency. Alternatively, a single threadblock can be employed to perform a certain algorithm, synchronization within a threadblock is possible since it runs the same piece of physical hardware. Kernels in different streams run parallel to each other. In this case, \textit{events} can be used to halt one stream until a certain kernel in another stream has finished processing.

\subsection{Continuous real-time processing requirements}
The real-time \gls{GPU} receiver consists of a \SI{1}{\GHz} photodiode connected to a \SI{12}{bit} \SI{4}{GSa\per\s} \gls{ADC}. Digitized samples are copied in buffers from the \gls{ADC} to the \gls{GPU} where they are processed in a highly parallel manner. Each buffer contains 2\textsuperscript{22} samples, which takes \SI{1.049}{\ms} at \SI{4}{GSa\per\second}. In our implementation, each buffer is assigned its own processing stream and any dependencies to ensure data continuity are handled by events. For real-time processing, the buffers need to be processed as fast or faster than they are created by the \gls{ADC} in order to avoid data loss. As such, the average buffer processing time needs to be lower than \SI{1.049}{\ms} times the number of streams employed. Therefore, buffer processing times can be relaxed by increasing the number of parallel streams at the expense of increased latency.

\begin{figure*}
\includegraphics[width=\linewidth]{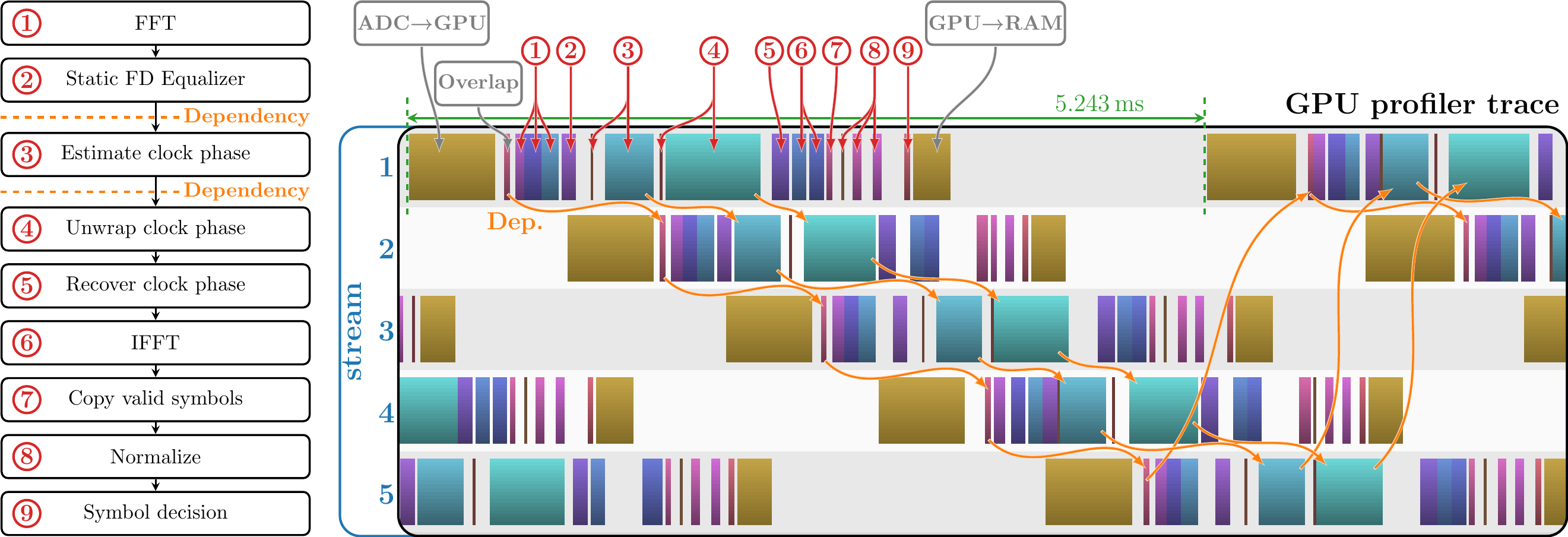}
\caption{\gls{IMDD} \gls{GPU} signal processing chain including an annotated \gls{GPU} profiler trace, detailing the \textit{Signal Processing} block of \cref{fig:rxarch} for \PAM{N} signals.}
\label{fig:imdd}
\end{figure*}

\subsection{GPU signal processing structure}
\label{ssec:rxarch_gpuchain}

\cref{fig:rxarch} shows the structure of receiver, the tasks performed by the \gls{GPU}, and how those are controlled by the \gls{CPU}. The \gls{ADC} is controlled by an \gls{API}, provided by the manufacturer which also manages the data transfer to the \gls{GPU}. A second program, written by the authors, controls the \gls{API} and launches signal processing kernels.

\begin{itemize}
	\item \textbf{Step 1:} The \gls{ADC} digitizes the analog signal into 12-bit digital samples at either 2 (\gls{IMDD}) or 4 (\gls{KK}) samples per symbol and temporarily stores them in \gls{ADC} memory. The \gls{API} initiates the transfer of a buffer containing 2\textsuperscript{22} samples from \gls{ADC} memory to \gls{GPU} memory using \gls{DMA}, provided a free \gls{GPU} buffer is available for the \gls{API} to use. This is marked as Step 1 in \cref{fig:rxarch}. Each buffer is assigned its own stream and \gls{DSP} kernels are added to that stream to process the data.

	\item \textbf{Step 2:} Control over the \gls{GPU} buffer which now contains the digitized signal is handed over to the control program written by the authors.

	\item \textbf{Step 3:} For continuous real-time data processing certain overlap between buffers is required, an overlap kernel is used for this. These overlap kernels need to be executed in order and events ensure an overlap kernel cannot start processing until its predecessor is finished. This is shown in \cref{fig:rxarch} as Step 3 and marked as \textit{Dependency}.

	\item \textbf{Step 4:} The 2\textsuperscript{22} samples in a buffer are subdivided into 8192 blocks of 512 samples for \gls{FD} processing as depicted in \cref{fig:dataarch}. \Gls{FD} processing requires one block of overlap between buffers for data continuity. An overlap kernel handles this by prepending a block to the current buffer which was stored elsewhere in memory. Afterwards, it copies the last block of its buffer to memory for the next overlap kernel to use. Also, the overlap kernel converts the data from 12-bit unsigned integers to 32-bit floats.

	\item \textbf{Step 5:} This step contains the actual \gls{DSP} chain which uses both \gls{TD} and 100\% overlap-save \gls{FD} processing. This block uses floating point samples as input and produces decoded bits as output. A detailed description can be found in \cref{sec:imdd,sec:kk} for \pam{N} and \qam{N} signals, respectively.

	\item \textbf{Step 6:} After processing, the decoded bits are copied to \gls{RAM} and control over the buffer is handed back to the \gls{API}.
\end{itemize}

\section{\gls{IMDD} \gls{GPU} Signal Processing Chain}
\label{sec:imdd}

\cref{fig:imdd} lists the \gls{DSP} chain for \gls{IMDD} \pam{N} signals. It consists of 9 steps, each of which executed by one or more kernels. To support real-time operation, five parallel streams are used as shown in the annotated profiler trace in \cref{fig:imdd}. Dependencies between algorithms running in parallel streams are handled through events which are marked as \textit{Dependency} in the \gls{DSP} chain and shown by the orange arrows in the profiler trace.

\subsection{Step 1 and 2: \gls{FFT} and Static Equalization}
\label{ssec:imdd-fdeq}
The \gls{IMDD} signal processing chain starts after overlap copying. A 100\% overlap-save 1024-point \gls{FFT} at 2 samples-per-symbol is performed using a readily-available highly-parallel \gls{GPU} \gls{FFT} implementation. This splits the 2\textsuperscript{22} samples in the buffer into 8192 blocks of 1024 samples, of which 512 are \textit{valid} due to 100\% overlap-save. Secondly, static \gls{FD} equalization is performed to compensate for receiver bandwidth impairments using a pre-computed \gls{FIR} filter. This filter optimized offline in \gls{TD} using 503 taps, converted to a 1024-point \gls{FD} version, and uploaded to the \gls{GPU}. The number of taps was limited to 503 to prevent introduction of \gls{ISI} through the cyclic nature of the 1024-point \gls{FFT}.

To fully appreciate the parallel nature of this processing, we need to look at the number of independent threads working in this one kernel alone. \gls{FD} equalization requires 512 complex multiplications to be performed for each of the 8192 blocks, Hermitian symmetry allows for the omission of half of the spectrum. To this end, 2\textsuperscript{21} threads are launched, each operating on 2 complex samples (4 32-bit floats) at a time. 128-bit vector loads/stores allow for the 4 floats to be loaded/stored using just a single instruction, increasing memory throughput. These 2\textsuperscript{21} threads can be performed in parallel, exploiting the massive parallel capabilities of the \gls{GPU}. \cref{fig:imdd} shows that during the execution of this kernel in stream 1, marked as step~2, is performed in parallel with a \gls{ADC}-to-\gls{GPU} copy in stream 2 and other stages of the signal processing of other buffers in streams 4 and 5. Therefore, parallelization is not only exploited \textit{within} kernels acting on a buffer, but also \textit{between} streams operating on different buffers.

\subsection{Step 3 and 4: Clock-phase Estimation and Unwrapping}
\label{ssec:imdd-clockest}
Clock-phase estimation is performed block-wise in \gls{FD} after static equalization using a technique introduced in \cite{FDClockPhase}. This provides an estimate clock-phase for each block of samples. To improve noise tolerance, these estimates are averaged over 105 blocks. This requires the 52 previous and 52 \textit{future} clock-phase estimates to be known as well. The causality issue of the future estimates is resolved through increased buffering in the \textit{overlap} kernel before actual signal processing starts. The dependency on previous estimates requires the clock-phase estimation of the previous buffer to be completed before the averaging and unwrapping step of the current buffer can be allowed to start. To this end, \textit{events} are used to signal when clock-phase estimation is completed, allowing for the current processing to wait until the previous has completed. Note that only the estimation step has this dependency, the remainder of the signal processing can occur in parallel. The events resolving these dependencies are shown by orange arrows in \cref{fig:imdd}.

The clock-phase estimates are restricted to $2\pi$. Hence, averaging is performed through vector addition in complex space and subsequent phase unwrapping is required. It is denoted as step 4 in \cref{fig:imdd}. The phase unwrapping kernel checks whether the current averaged clock-phase differs more than $\pi$ from the previous. This sequential algorithm is hard to parallelize. To some extent this is done through inter-thread communication using warp-level shuffles. This requires some significant processing time. However, the unwrapping algorithm uses a single warp of 32 threads and leaves much of the \gls{GPU} processing power unused, which can be used by other kernels running in different streams. For example, during the phase unwrapping in stream 1, stream 2 performs an \gls{FFT}, \gls{FD} equalization, and clock-phase estimation, stream 3 performs a \gls{ADC}-to-\gls{GPU} memory copy, stream 4 is idle, and stream 5 performs normalization, symbol decision, and a \gls{GPU}-to-\gls{RAM} copy. Therefore, phase unwrapping does not take up significant amount of \textit{resources}, even though it takes up significant amount of \textit{time}.

\begin{figure}
\includegraphics[width=\linewidth]{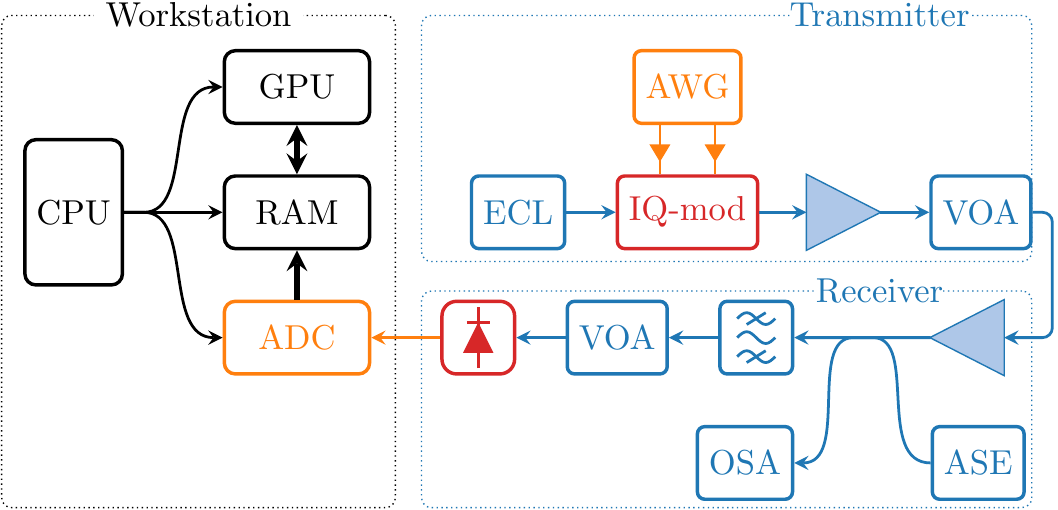}
\caption{Experimental noise-loading setup for back-to-back evaluation of the real-time receiver.}
\label{fig:expb2b}
\end{figure}

\subsection{Step 5-9: Clock Recovery, IFFT, Normalization, and symbol decision}
Clock recovery is performed by correcting for the unwrapped clock-phase in \gls{FD}. After the 1024-point \gls{IFFT}, 256 \textit{valid} symbols need to be extracted for further processing. In the presence of clock-frequency offset, every now and then, either more or fewer symbols may need to be extracted from a block to keep the unwrapped clock-phase within bounds. This is performed in step~7 of \cref{fig:imdd}, which converts the fixed rate sample input to a variable rate symbol output. Then, buffer-wise normalization is performed using three kernels: initialization, estimation of the DC-offset, and estimation of the amplitude. In the symbol decision kernel, the DC-offset and amplitude are corrected for and \PAM{N} symbols are decoded into bits. Decision thresholds are optimized offline beforehand and uploaded to the \gls{GPU}.

\subsection{Experimental setup for back-to-back evaluation of \pam{N}}
\label{ssec:pamsetup}
\cref{fig:expb2b} shows a diagram of the experimental setup for back-to-back characterization of the real-time receiver. At the transmitter, the lightwave from a \SI{500}{kHz} linewidth \gls{ECL} centered at \SI{1542.92}{nm} is modulated using a single-polarization \gls{IQM}. Electrical driving signals for the \gls{IQM} are provided by a 2-channel \gls{AWG} operating at \SI{12}{Gs\per \second} amplified by RF-amplifiers, whilst bias-tees and voltage sources control the bias of the modulator arms. \PAM{N} signals are modulated by biasing one of the \gls{IQM}-arms to mid-point and driving it with a baseband 2~Gbaud 50\% roll-off \gls{RRC} pulse-shaped signal. 

The receiver consists of an \gls{EDFA} pre-amplifier followed by a 0.04~nm \gls{BPF}. In addition, a noise-loading setup is included with an \gls{ASE} source and an \gls{OSA} through a 2$\times$2 coupler. A \gls{VOA} is used to control the power at the input of a \gls{PD} with a \SI{3}{dB} cut-off frequency of \SI{1}{GHz}. The electrical \gls{PD} output is directed to the \gls{ADC} for processing.

\subsection{Experimental results}
\begin{figure}
\includegraphics[width=\linewidth]{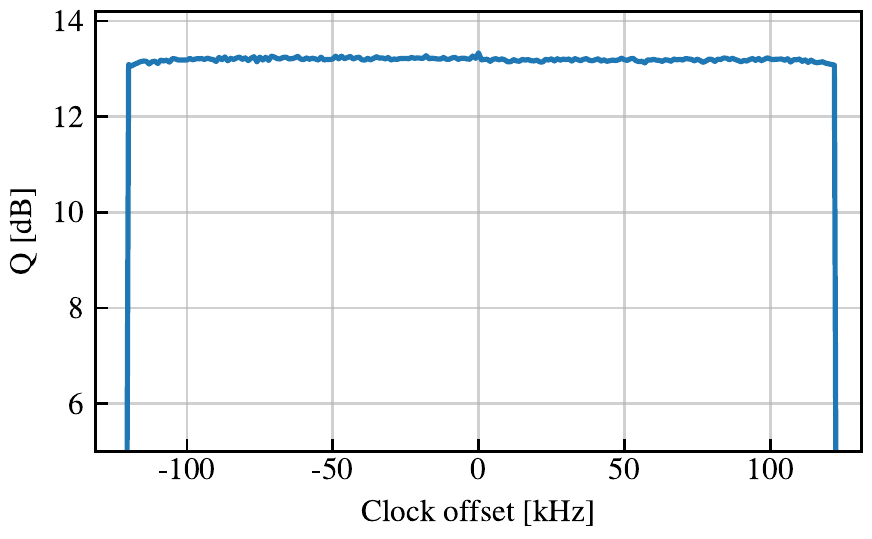}
\caption{Q-factor versus clock-frequency offset. The clock-recovery algorithms allows for stable performance across a wide range of clock-frequency offsets.}
\label{fig:clockoffset}
\end{figure}

\begin{figure}
\includegraphics[width=\linewidth]{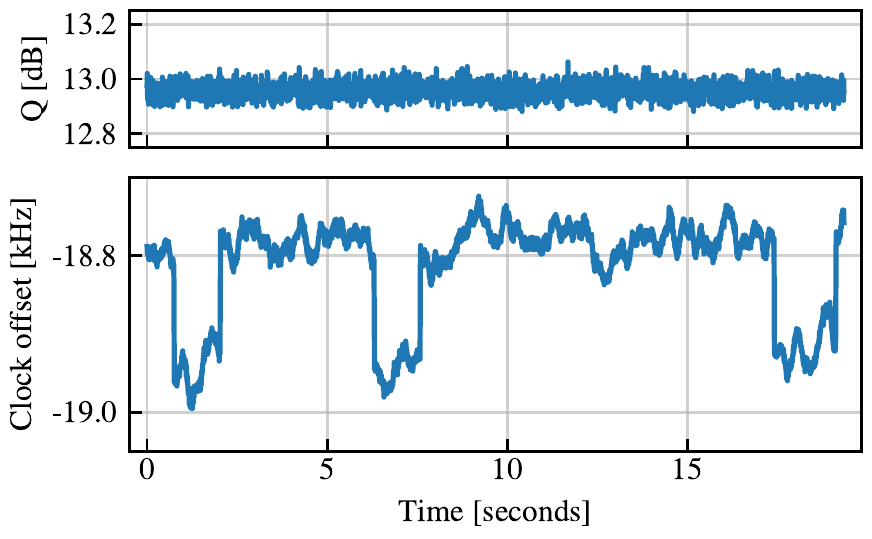}
\caption{Clock-frequency offset and Q-factor versus time when using free-running clocks. The clock-recovery algorithm allows for stable performance even when clock-frequency offset changes rapidly.}
\label{fig:clockphase}
\end{figure}

\begin{figure}
\includegraphics[width=\linewidth]{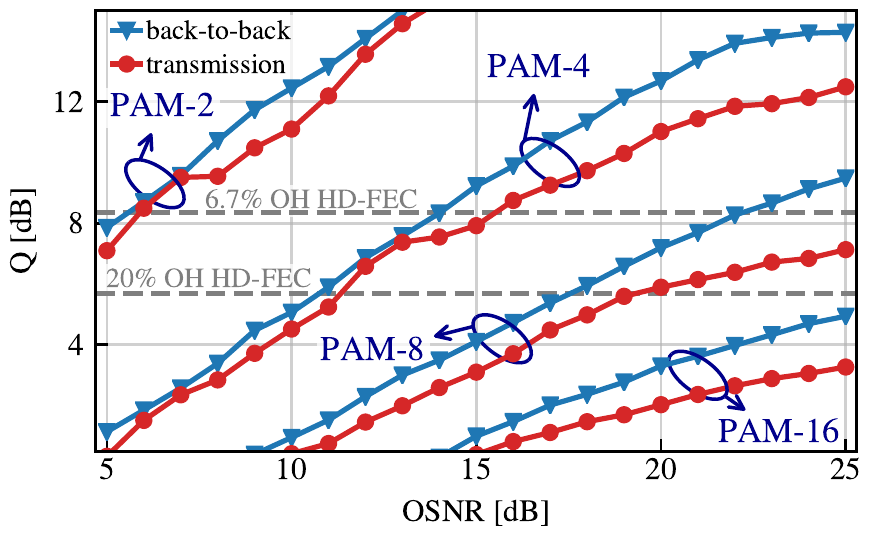}
\caption{Q-factor versus OSNR for \pam{N} signals, both back-to-back and after transmission over the field trial network.}
\label{fig:pamresults}
\end{figure}

\begin{figure*}
%%%%% KK figure, belongs to the next section%%%%%
\includegraphics[width=\linewidth]{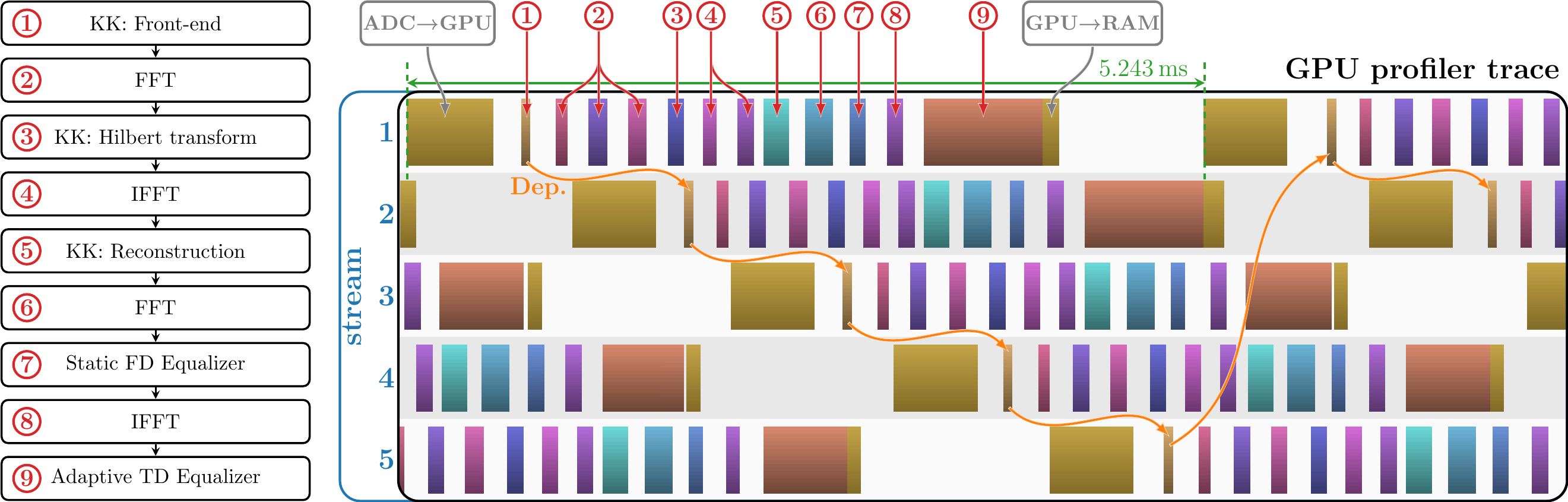}
\caption{\gls{KK} \gls{GPU} Signal Processing Chain including an annotated \gls{GPU} profiler trace, detailing the \textit{Signal Processing} block of \cref{fig:rxarch} for \QAM{N} signals.}
\label{fig:kk}
\end{figure*}

The performance of clock-recovery is evaluated using \PAM{4} and \PAM{8} signals in back-to-back transmission. \cref{fig:clockoffset} shows the Q-factor versus the clock-frequency offset between transmitter and receiver clock when transmitting \SI{2}{GBaud} \PAM{4} signals. An attenuator limited the power into the photodiode to \SI{-10}{dBm} to introduce enough noise and thus bit errors to properly evaluate performance when changing the clock-frequency offset. Performance is stable for a wide range of offsets, showing the resiliance of the implemented algorithms. Performance drops off very rapidly when an offset of \SI{122}{kHz} (\SI{30.5}{ppm}) or more is applied, which can be attributed an implementation choice to use an 8-bit integer to keep track of number of symbols added or removed throughout the buffer. A change to a 16- or even 32-bit number would greatly increase clock-frequency offset tolerance, but was deemed unnecessary.

\cref{fig:clockphase} shows clock-frequency offset and Q-factor over time for \SI{2}{GBaud} \PAM{8} signals at \SI{0}{dBm} input power when using free-running clocks. The transmitter \gls{DAC} is driven by a laboratory-grade tone-generator whilst the \gls{ADC} uses its own internal clock source. Even when the clock-frequency offset experiences rapid changes as shown in \cref{fig:clockphase}, the Q-factor remains constant, demonstrating that the clock-recovery algorithm is able to cope with these rapid transitions. Since the \gls{ADC} manufacturer advises against the use of the internal clock, the authors consider this a worst-case test. For the remainder of this work, the \gls{ADC} received a high-quality clock-signal from a laboratory-grade tone-generator.

\cref{fig:pamresults} shows the Q-factor as a function of \gls{OSNR} for \pam{2}, \pam{4}, \pam{8}, and \pam{16}. In back-to-back, performance reaches the \SI{8.4}{dB} Q-factor threshold for 6.7\% overhead \gls{HDFEC}\cite{agrell_information-theoretic_2018} at \SI{5.6}{dB}, \SI{14.0}{dB}, and \SI{22.2}{dB} for \pam{2}, \pam{4}, and \pam{8}, respectively. \pam{16} can be decoded in real-time using the \gls{GPU} \gls{DSP}, but signal quality is not sufficient to reach the threshold for either 6.7\% or 20\%\cite{agrell_information-theoretic_2018} overhead \gls{HDFEC}. Most likely this is due to severe low-pass filtering of the signal by the receiver components. The \SI{2}{GBaud} signal with 50\% \gls{RRC} roll-off uses \SI{1.5}{GHz} of electrical bandwidth, whilst the \SI{3}{dB} bandwidth of the photodiode and \gls{ADC} are both \SI{1}{GHz}. The static equalizer (see \cref{sec:imdd} and \cref{fig:imdd}, step 2) can boost the attenuated higher frequencies, but only at the cost of amplifying noise. Future \glspl{ADC} (PCIe Gen 4) offer greater bandwidth and sampling rate, facilitating greater baud and data rates. Proprietary interfaces such as NVIDIA NVLink \cite{NvidiaNVLink} can support a further tenfold increase.

\section{\gls{KK} \gls{GPU} Implementation and Evaluation}
\label{sec:kk}

We chose to implement KK field reconstruction to showcase GPU excellence in handling large FFTs and exploiting its enhanced capability for frequency-domain signal processing. \cref{fig:kk} shows the \gls{DSP} chain for \gls{KK} \qam{N} signals subdivided in 9 steps, each of which an algorithm performed by one or more \textit{kernels} as described in this section. Five parallel streams are used as shown in the profiler trace. Dependencies between streams are marked as \textit{Dependency} and annotated with orange arrows in the profiler trace.

\subsection{Step 1: Overlap and KK Front-end}
The \gls{KK} Front-end containing the square root and logarithm operations are incorporated into the overlap kernel to limit \gls{GPU} memory access and thus improve performance. The overlap part of this kernel, including the dependency handling via \textit{events}, see \cref{ssec:rxarch_gpuchain} and \cref{fig:rxarch}. 

Since the digitizer used in this experiment was AC-coupled, no DC-terms are measured, hampering \gls{KK} field reconstruction. Therefore, an offline-optimized static DC-offset is added to the signal \cite{Luis:20} after the data are converted from 12-bit unsigned integers to 32-bit floats. Subsequently, a conventional \gls{KK} front-end\cite{mecozzi_kramers_2016} performs the square root, to retrieve the signal amplitude, and logarithm, required for phase reconstruction, operations at 4 samples-per-symbol.

\subsection{Step 2-5: Hilbert transform and KK field reconstruction}
A 100\% overlap-save 1024-point real-to-complex \gls{FFT} is used to convert the samples pre-processed for phase-retrieval by the \gls{KK} front-end to frequency domain, dividing the 2\textsuperscript{22} samples in the buffer in 8192 blocks of 1024 samples of which, because of 100\% overlap-save, 512 are \textit{valid}. The Hilbert transform is performed in \gls{FD} before a complex-to-complex \gls{IFFT} converts back to \gls{TD}. Now, the \gls{KK} field reconstruction\cite{mecozzi_kramers_2016} combines the previously retrieved signal amplitude with the phase recovered through the logarithm and Hilbert transform. The recovered optical field is downshifted to DC for further processing.

\subsection{Step 6-8: FD static equalization}
After a 1024-point complex-to-complex \gls{FFT}, the recovered signal is filtered in \gls{FD} by a static 203-tap \gls{FIR} filter, which is optimized offline beforehand and uploaded to \gls{GPU} memory. This static equalizer compensates for receiver bandwidth impairments and performs matched filtering for the \gls{RRC} \QAM{N} signals. A 512-point \gls{IFFT} both converts the signal to \gls{TD} and downsamples it to 2 samples-per-symbol. 

\begin{figure}
\includegraphics[width=\linewidth]{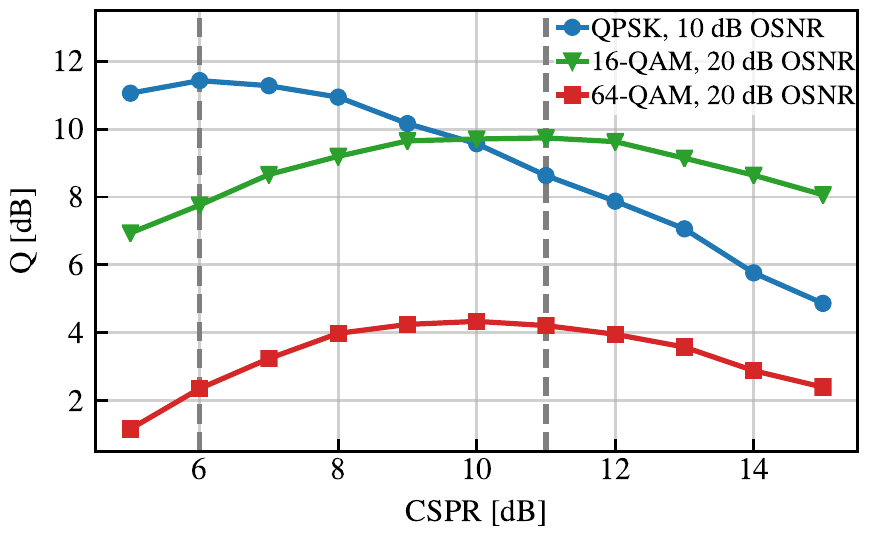}
\caption{CSPR optimization shows a trade-off between reconstruction errors at low CSPR and increased noise at high CSPR. \SI{6}{dB} is chosen for \qam{4} and \SI{11}{dB} for \qam{16/64}.}
\label{fig:csprresults}
\end{figure}

\subsection{Step 9: TD adaptive equalization and symbol decision}
Clock-phase and symbol-phase recovery, transmitter IQ-imbalance compensation, and symbol decision and demapping are performed by a 4-tap adaptive widely-linear\cite{Silva_WidelyLinear} \gls{TD} \gls{DDLMS} equalizer. Note that in contrast to the \PAM{N} signals of \cref{sec:imdd}, a \SI{10}{MHz} reference clock was shared by transmitter and receiver, so the equalizer only needs to handle relatively small clock-phase and symbol-phase fluctuations, for example due to changing conditions in the field-deployed fiber. During equalization, the decisions made by the equalizer are demapped and stored in \gls{GPU} memory to be sent to \gls{RAM} after this kernel is finished.

Four taps was deemed sufficient and has the benefit of exploiting 128-bit parallel data access through vector load/store instructions as explained in \cref{ssec:imdd-fdeq}. Furthermore, warp-level shuffles are used to further optimize this \gls{TD} adaptive equalizer kernel which is serial in nature. One might conclude based on the \gls{GPU} profiler trace in \cref{fig:kk} that this kernel uses a lot of resources since it uses a lot of time. However, this is not correct. A relatively low amount of \gls{GPU} parallel processing units are used for execution of this kernel. Therefore, this kernel does not take up significant amount of \textit{resources} even though it takes up significant amount of \textit{time}, similar to the clock-phase unwrapping kernel discussed in \cref{ssec:imdd-clockest}. The unused parallel processing units can be used by other parallel processing streams, see \cref{fig:kk}.

\subsection{Back-to-back evaluation of N-QAM signals}
\label{ssec:kksetup}
\Gls{KK} \QAM{N} signals are generated using the same setup as \PAM{N} signals explained in \cref{ssec:pamsetup,fig:expb2b}. However, the \gls{IQM} is operated at the minimum optical output bias point, whilst the \gls{AWG} produces baseband 1~GBaud \QAM{N} signals with 1\% roll-off \gls{RRC} pulse shaping combined with a digitally-introduced carrier tone at 0.547~GHz. The tone power can be chosen to produce the desired \gls{CSPR}. Note that the \qam{N} required optical bandwidth is half of \pam{N}, but the required electrical bandwidth is identical.

\begin{figure}
\includegraphics[width=\linewidth]{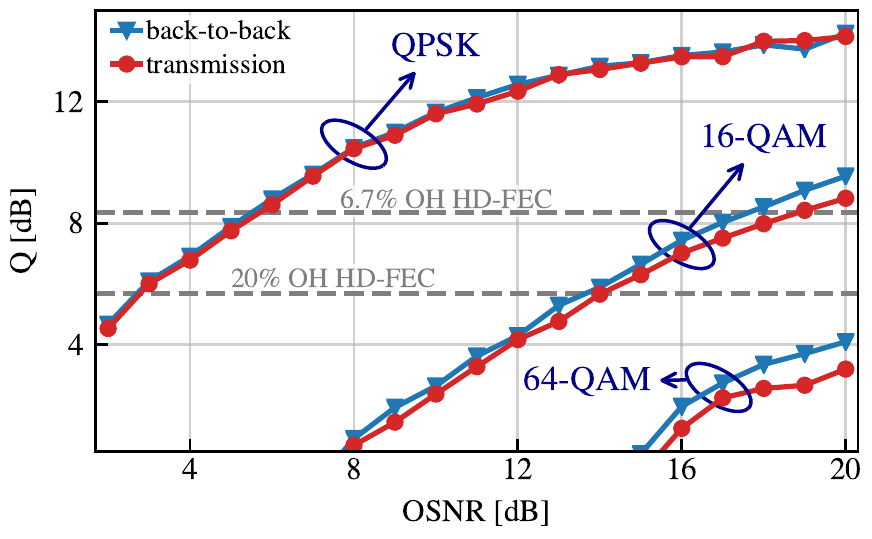}
\caption{Q-factor versus OSNR for Kramers-Kronig \qam{N} signals, both back-to-back and after field trial transmission.}
\label{fig:kkresults}
\end{figure}

\Gls{CSPR} optimization is important for \gls{KK} \qam{N} signals since it directly influences the accuracy of signal reconstruction and \gls{OSNR} performance. When employing high carrier power, \gls{SSBI} is lower and signal reconstruction through the \gls{KK} algorithm is better, thus improving signal quality after receiver \gls{DSP}. However, higher carrier power leads to lower signal power for the same combined power. Therefore, signal quality degrades in the higher \gls{CSPR} region, as can be seen in \cref{fig:csprresults}. The choice of \gls{CSPR} is essentially a trade-off between increased reconstruction error at lower \glspl{CSPR} versus increased noise at higher \glspl{CSPR}. Moreover, the optimal choice also depends on modulation cardinality, since high-cardinality modulation formats such as \qam{64} suffer more from reconstruction errors than \qam{4}. For simplicity of measurement, the \gls{CSPR} is optimized at only one specific value for \gls{OSNR}, \SI{10}{dB} for \qam{4} and \SI{20}{dB} for \qam{16} and \qam{64}. A \gls{CSPR} of \SI{6}{dB} is chosen for \qam{4} whilst \SI{11}{dB} is employed for \qam{16} and \qam{64} throughout this work.

\cref{fig:kkresults} shows the Q-factor as a function of \gls{OSNR} for \qam{4, 16, and 64}. \qam{4} reaches the 6.7\% overhead \gls{HDFEC} threshold\cite{agrell_information-theoretic_2018} at \SI{5.5}{dB} \gls{OSNR}, whilst \qam{16} requires an \gls{OSNR} of \SI{17.6}{dB}. \qam{64} signals were received and processed in real time, however, performance was not sufficient to reach either the 6.7\% of 20\% overhead \gls{HDFEC} threshold\cite{agrell_information-theoretic_2018}.

% The newlines above are important because otherwise Figure 12 is bumped to page 9 instead of the top of page 8 where it should be.
% \input{sections/expsetup}
\begin{figure*}
\includegraphics[width=\linewidth]{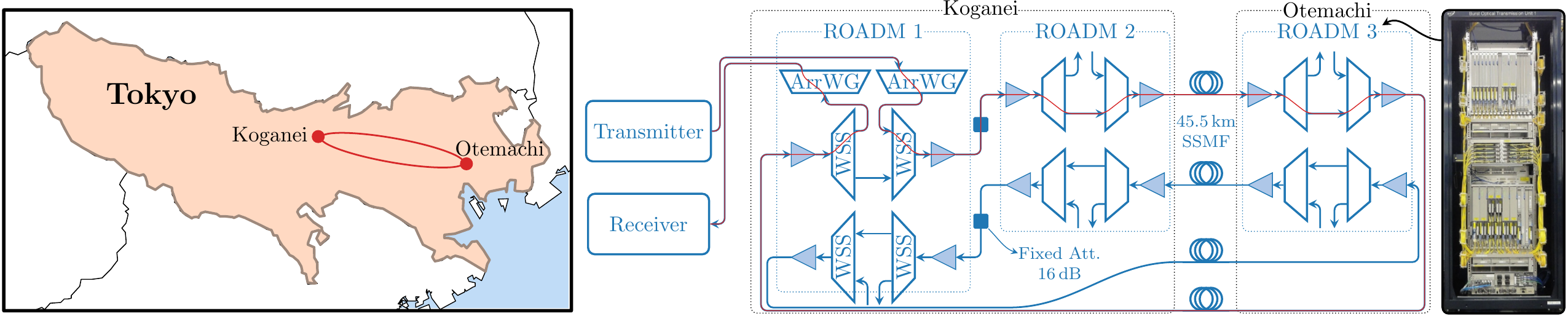}
\caption{Experimental setup using field-deployed fiber between Koganei and Otemachi, Tokyo. Transmitter and receiver structure are detailed in \cref{fig:expb2b}.}
\label{fig:expsetup}
\end{figure*}%
\section{Experimental field trial}
\label{sec:results}
\subsection{Experimental setup}
The same transmitter and receiver architecture used for back-to-back characterization, see \cref{fig:expb2b,ssec:pamsetup,ssec:kksetup}, is also used to generate and receive the signals in the field trial scenario. The signal launch power is set by an \gls{EDFA} followed by a \gls{VOA}. The transmission network shown in \cref{fig:expsetup} consists of a bidirectional ring with 3 commercial \glspl{ROADM}. Two \glspl{ROADM} are installed in the same location in Koganei, Tokyo. The link between these \glspl{ROADM} is relatively short and its loss was set to \SI{16}{dB} using fixed attenuators. Both \glspl{ROADM} are connected to a commercial \gls{ROADM} in Otemachi, Tokyo by a \SI{45.5}{km}, 4-fiber link. The transmission loss, including optical distribution frames, is \SI{16.5}{dB}. 56\% of the fiber is installed in underground ducts and the remainder on areal paths and in the surface along railway tracks. The red line in \cref{fig:expsetup}) shows the signal path along the network, with a total transmission distance of \SI{91}{km}. Each \gls{ROADM} has two line sides, each consisting of \glspl{WSS} and optical amplifiers for add/drop and express connections. In addition, \glspl{ARRWG} were used for add and drop. \cref{fig:expsetup} shows a photograph of one of the commercial \glspl{ROADM}. 

\subsection{Transmission results of \pam{N} signals}

\begin{figure}
	\centering
	\begin{subfigure}{0.22\linewidth}
		\centering
		\includegraphics[width=\linewidth]{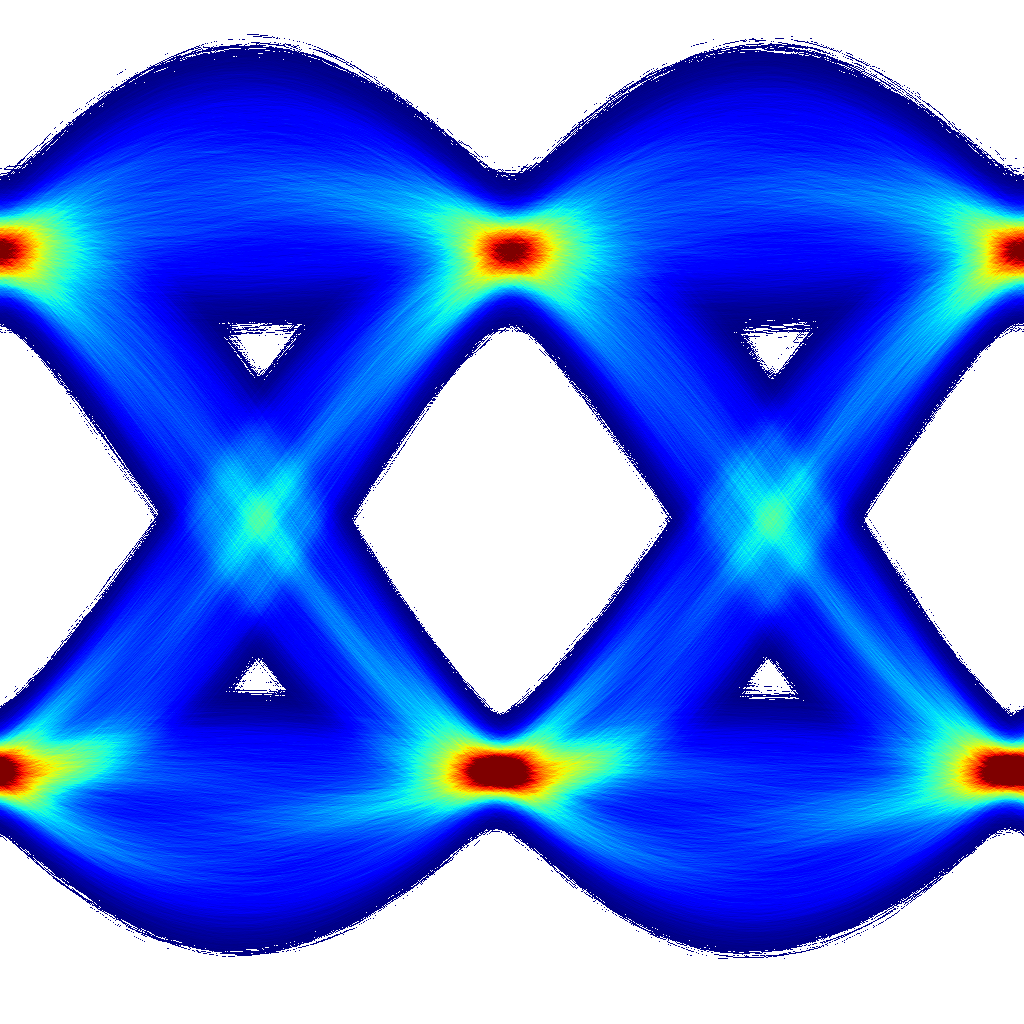}
		\caption{PAM-2}
		\label{fig:eyepam2}
	\end{subfigure}%
	\hfill%
	\begin{subfigure}{0.22\linewidth}
		\centering
		\includegraphics[width=\linewidth]{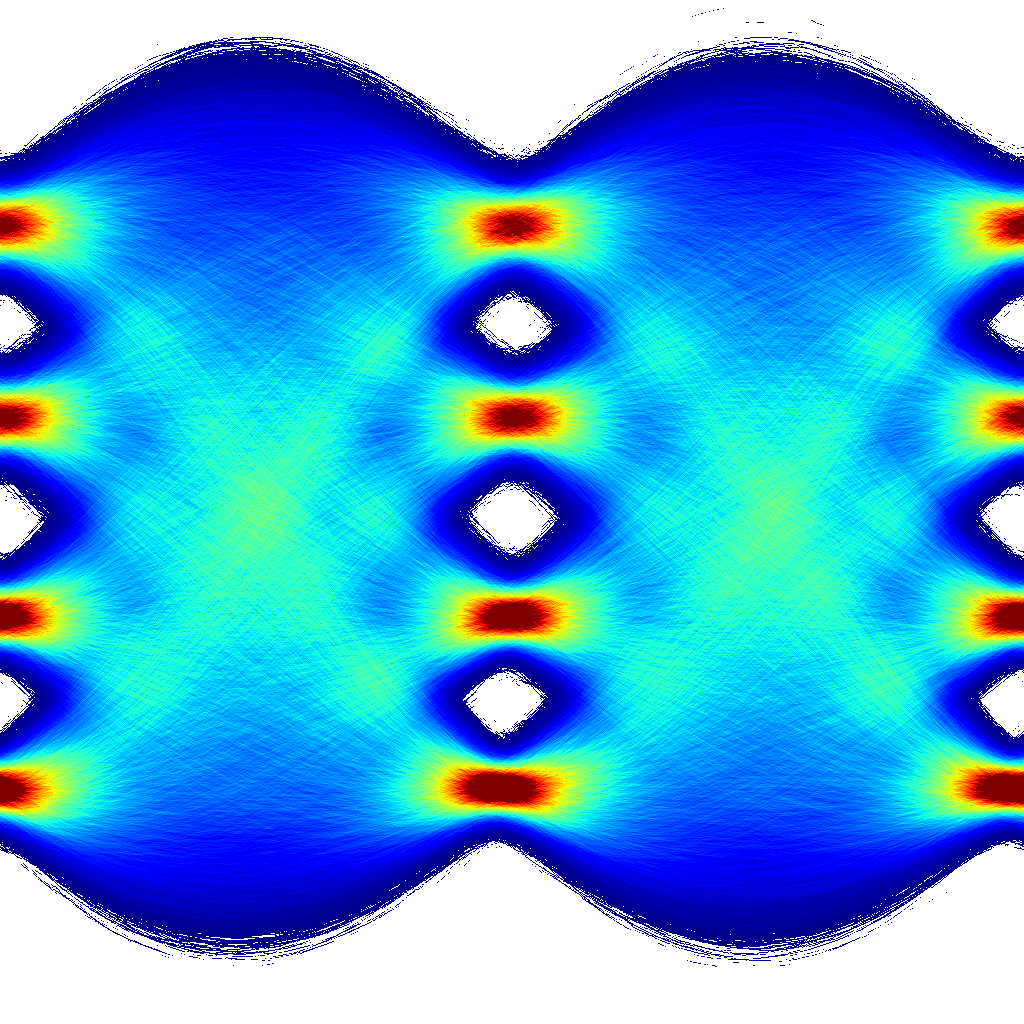}
		\caption{PAM-4}
		\label{fig:eyepam4}
	\end{subfigure}%
	\hfill%
	\begin{subfigure}{0.22\linewidth}
		\centering
		\includegraphics[width=\linewidth]{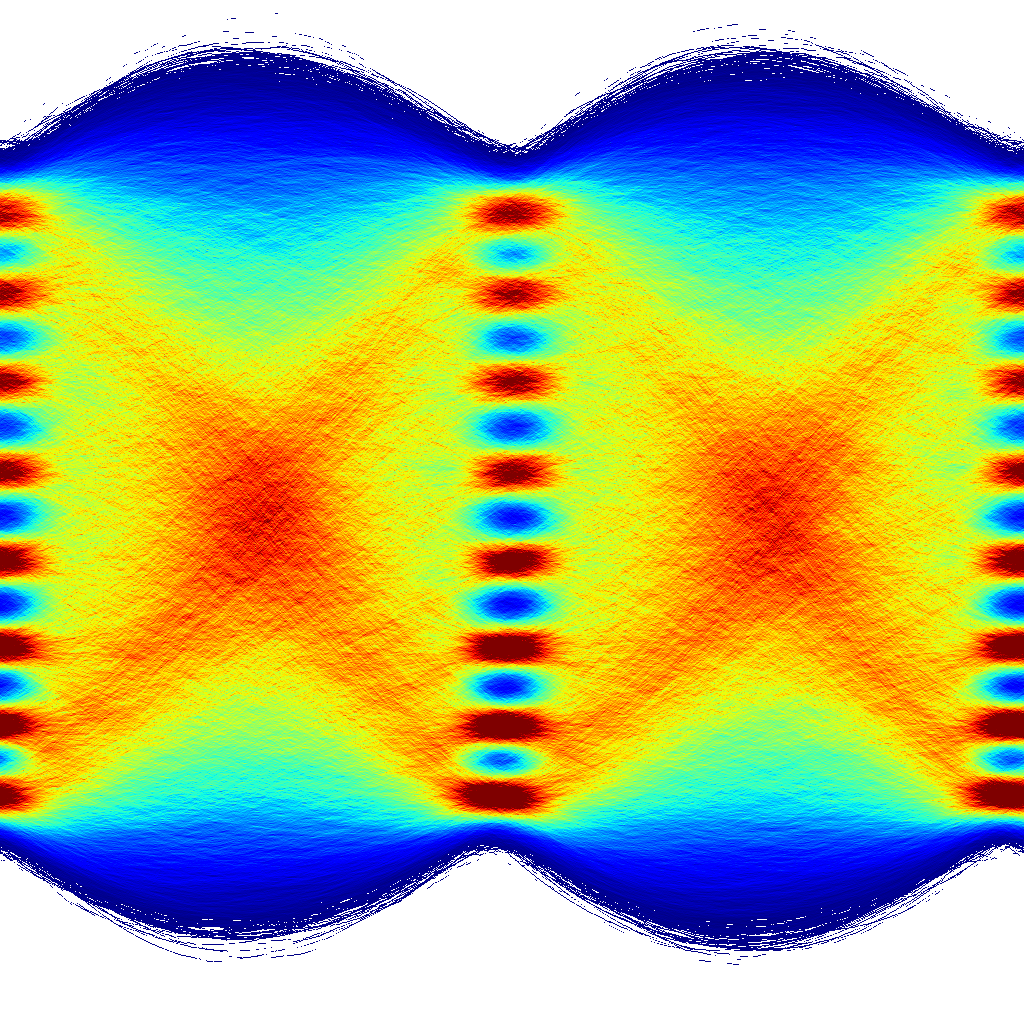}
		\caption{PAM-8}
		\label{fig:eyepam8}
	\end{subfigure}%
	\hfill%
	\begin{subfigure}{0.22\linewidth}
		\centering
		\includegraphics[width=\linewidth]{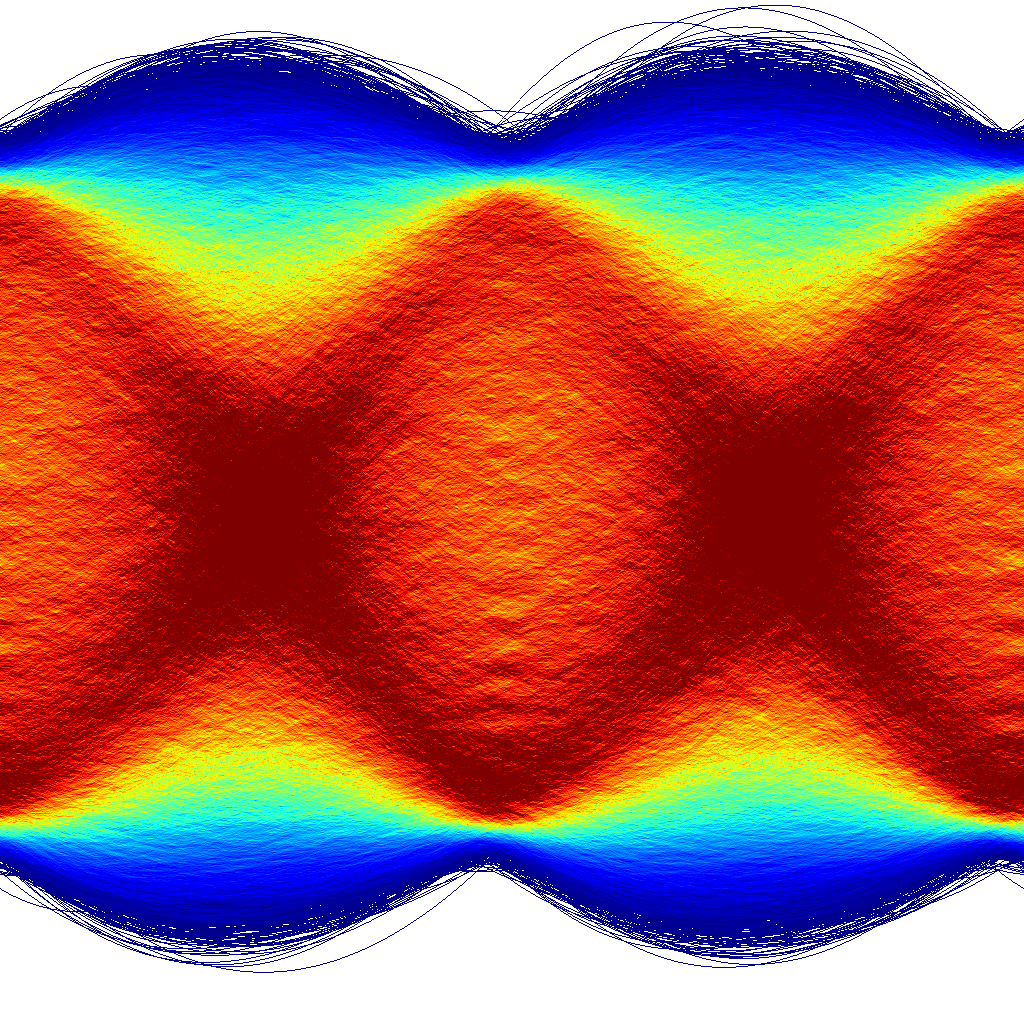}
		\caption{PAM-16}
		\label{fig:eyepam16}
	\end{subfigure}%
	\caption{Eye diagrams for \pam{N} transmission over the field trial network without noise loading.}
	\label{fig:eyespam}
\end{figure}

\cref{fig:pamresults} shows the Q-factor as a function of \gls{OSNR} for \pam{2}, \pam{4}, \pam{8}, and \pam{16} for back-to-back and after transmission through the field trial network. An \gls{OSNR} penalty increasing with modulation cardinality, is observed. The penalty at the 6.7\% overhead \gls{HDFEC} threshold is \SI{0.4}{dB} and \SI{1.5}{dB} for \pam{2} and \pam{4}, respectively. After transmission through the field trial network, \pam{8} cannot be recovered using a 6.7\% overhead \gls{HDFEC}, but can when a 20\% overhead \gls{HDFEC} with a Q-factor threshold of \SI{5.7}{dB} is employed\cite{agrell_information-theoretic_2018}. Eye diagrams for \pam{N} transmission over the field trial network without noise loading are plotted in \cref{fig:eyespam}.

\subsection{Transmission results of N-QAM signals}

\begin{figure}
	\centering
	\begin{subfigure}{0.3\linewidth}
		\centering
		\includegraphics[width=\linewidth, trim=70 10 70 10, clip]{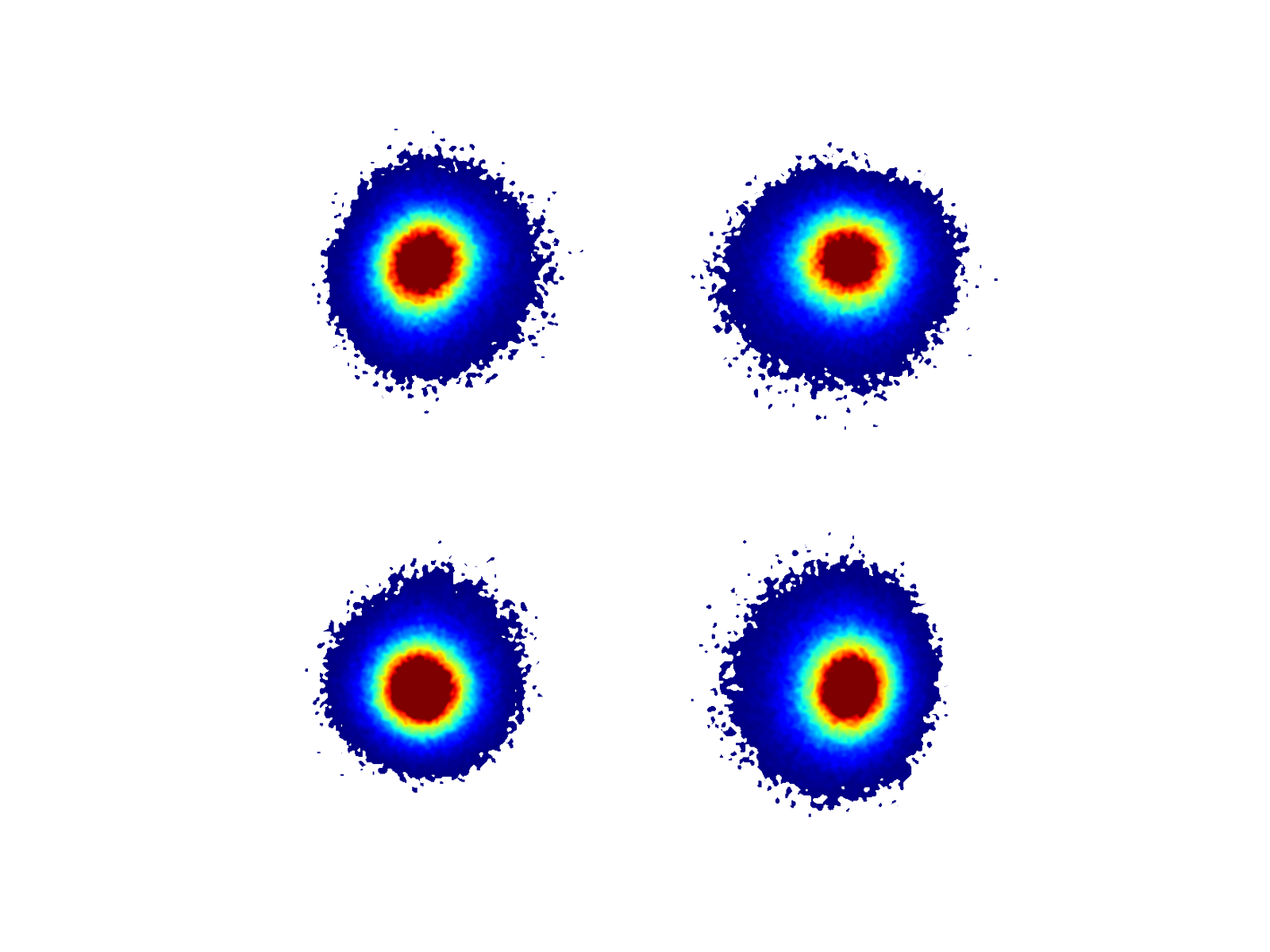}
		\caption{QPSK}
		\label{fig:const4qam}
	\end{subfigure}%
	\hfill%
	\begin{subfigure}{0.3\linewidth}
		\centering
		\includegraphics[width=\linewidth, trim=70 10 70 10, clip]{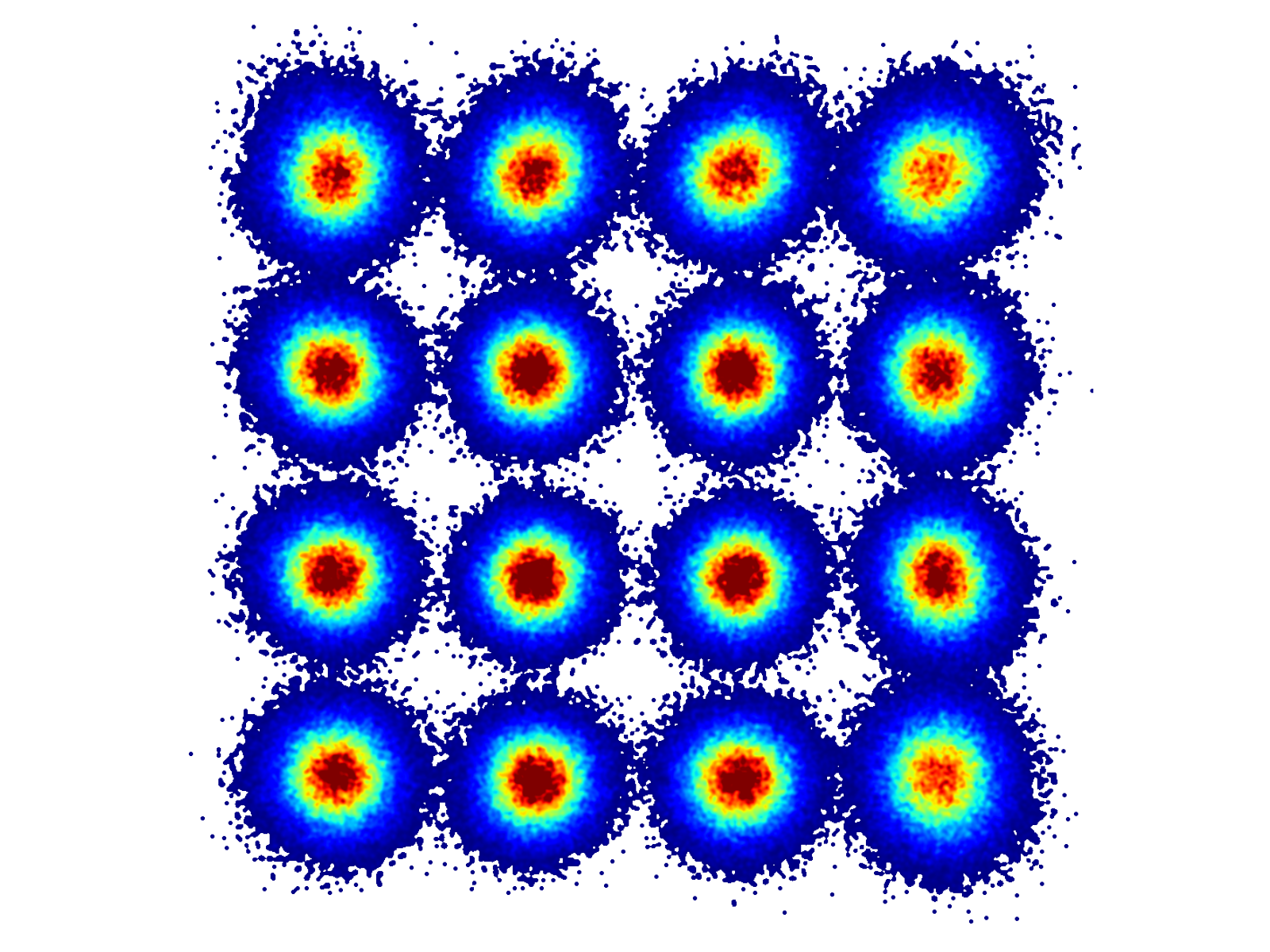}
		\caption{16-QAM}
		\label{fig:const16qam}
	\end{subfigure}%
	\hfill%
	\begin{subfigure}{0.3\linewidth}
		\centering
		\includegraphics[width=\linewidth, trim=70 10 70 10, clip]{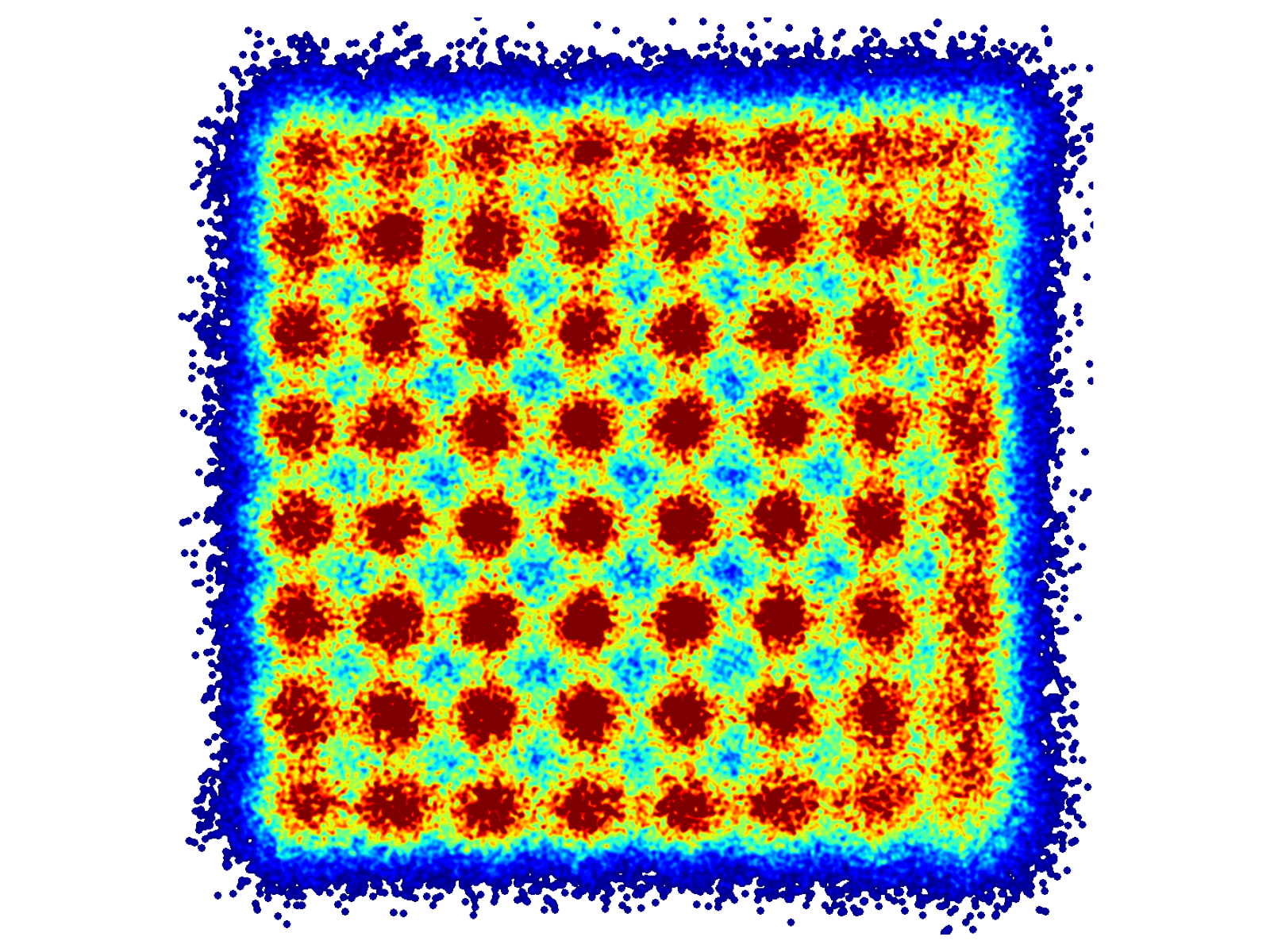}
		\caption{64-QAM}
		\label{fig:const64qam}
	\end{subfigure}%
	\caption{Constellation diagrams for KK \qam{N} transmission over the field trial transmission without noise loading.}
	\label{fig:constkk}
\end{figure}

\cref{fig:kkresults} shows the Q-factor as a function of \gls{OSNR} for \qam{4}, \qam{16}, \qam{64} for back-to-back and after transmission through the field trial network. A \SI{0.2}{dB} and \SI{1.2}{dB} \gls{OSNR} penalty is observed for \qam{4} and \SI{}, respectively. \qam{64} signals were received and processed in real time, however, performance was not sufficient to reach either the 6.7\% of 20\% overhead \gls{HDFEC} threshold\cite{agrell_information-theoretic_2018}. Constellation diagrams for these modulation formats at maximum available \gls{OSNR} after transmission over the field trial network are plotted in \cref{fig:constkk}.

\begin{figure}
\includegraphics[width=\linewidth]{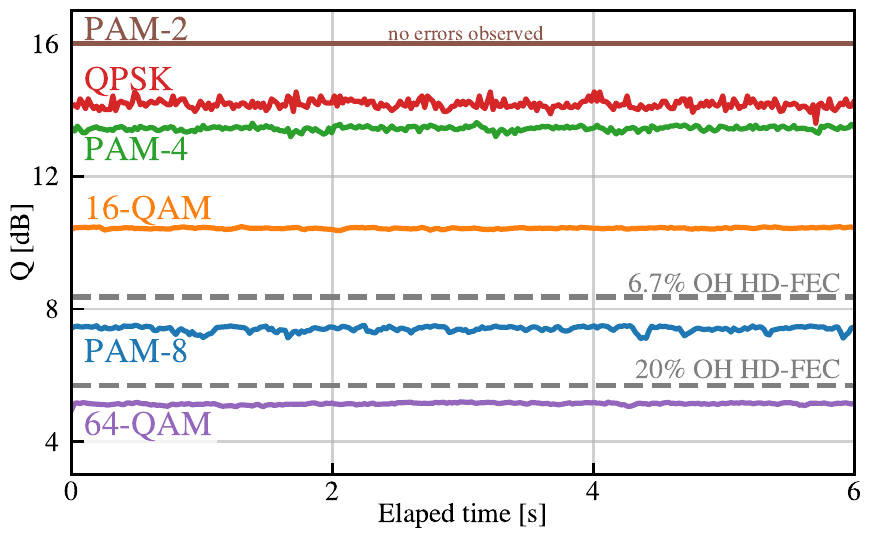}
\caption{Continuous real-time transmission traces for various modulation formats. Performance remains stable despite the varying environment of installed fiber.}
\label{fig:longtraces}
\end{figure}

\subsection{Continuous real-time transmission}
\cref{fig:longtraces} shows the short-term average Q-factor for six second long traces for various modulation formats. Within these six seconds, all transmitted symbols were received, processed, and recorded continuously, using the real-time \gls{GPU} algorithms detailed in the previous sections. During these six seconds, 6 billion symbols were received for \qam{N} signals and 12 billion symbols for \pam{N} signals. The Q-factor displayed in \cref{fig:longtraces} is estimated from the \gls{BER} in sections of \SI{21}{ms}. For all transmitted signals, we observe stable performance. No errors are observed while transmitting \pam{2}. \qam{4}, \pam{4}, and \qam{16} can be recovered using a 6.7\% overhead \gls{HDFEC} since the Q-factors are \SI{14.1}{dB}, \SI{13.4}{dB}, and \SI{10.4}{dB}, respectively. With a Q-factor of \SI{7.4}{dB}, \pam{8} cannot be recovered by the 6.7\% overhead \gls{HDFEC} but performance is sufficient for 20\% overhead error coding. \qam{64} is successfully transmitted, received, processed using the \gls{GPU}, and stored in \gls{RAM} in real time, but performance is not sufficient for \gls{HDFEC} algorithms considered.

\section{Conclusion}
\label{sec:conclusion}
A real-time, software-defined, multi-modulation-format, GPU-based receiver achitecture is introduced, detailed, and demonstrated to achieve stable real-time operation over a field-deployed metropolitan network. We show the potential for massive parallel processing provided by a GPU to recover directly-detected \pam{N} signals as well as \qam{N} signals with Kramers-Kronig coherent detection. \SI{2}{GBaud} optical signals using \pam{2}, \pam{4}, \pam{8}, and \pam{16} modulation, \SI{1}{GBaud} \qam{4}, \qam{16}, and \qam{64} modulation, are received and processed in real time by our flexible receiver architecture. \pam{2 and -4} and \qam{4- and 16} reach the Q-factor threshold for a 6.7\% overhead \gls{HDFEC} both in back-to-back and after transmission through the field-trial network. \pam{8} reaches this threshold in back-to-back, but no longer after transmission, although it can be received using a 20\% overhead \gls{HDFEC}. \pam{16} and \qam{64} are received and processed in real-time, but performance is not sufficient to reach either \gls{HDFEC} threshold. Continuous real-time transmission reveals stable performance despite the varying environment of installed fiber. These results show the potential of massive parallel processing provided by \glspl{GPU} for low-cost flexible optical links for a range of modulation formats.

% or


% \appendices

% \section{}

% \section*{Acknowledgment}


\ifCLASSOPTIONcaptionsoff
  \newpage
\fi



%\bibliographystyle{IEEEtran}

\printbibliography[notcategory=ignore]



% \end{thebibliography}


\begin{IEEEbiographynophoto}{Sjoerd van der Heide} (S'13)
was born in 's-Hertogenbosch, the Netherlands, in 1992. He received the B.Sc. and M.Sc. (cum laude) degrees in Electrical Engineering from Eindhoven University of Technology, the Netherlands, in 2015 and 2017, respectively. He is currently working towards a PhD degree at the High Capacity Optical Transmission Laboratory, Electro-Optical Communications group, at Eindhoven University of Technology. His research interests include space-division multiplexing and digital signal processing. He is the recipient of a student paper award at ECOC 2018 and a best paper award at OECC 2019.
\end{IEEEbiographynophoto}

\begin{IEEEbiographynophoto}{Ben Puttnam} (M’12) is a senior researcher in the Photonic Network System Laboratory at the National Institute of Information and Communications Technology (NICT) in Tokyo, Japan. He received the MPhys degree in Physics from the University of Manchester (UK) in 2000 and the PhD degree from University College London in 2008, working as a Switch Design Engineer for T-mobile (UK) in between. After short term visits to NICT, supported by JSPS and the Photonics group at Chalmers University, Göteborg, Sweden supported by the Ericsson research foundation he re-joined NICT in March 2010. His research interests are space-division multiplexing for optical transmission and optical signal processing.
\end{IEEEbiographynophoto}

\begin{IEEEbiographynophoto}{Georg Rademacher} (M’14, SM’20) received the Dipl.-Ing. and Dr.-Ing. degree in electrical engineering from Technische Universität Berlin, Germany, in 2011 and 2015, respectively. During his doctoral studies, he did internships at Bell Laboratories in Holmdel, USA and the National Institute of Information and Communications Technology (NICT) in Japan. In 2016 he joined the Photonic Network System Laboratory at NICT in Tokyo, Japan, where he is engaged in research on subsystems and systems for efficient high capacity optical transmission.
\end{IEEEbiographynophoto}

\begin{IEEEbiographynophoto}{Chigo Okonkwo} (M'09--SM'18)
was born in Wakefield, U.K., in 1979. He received the Ph.D. degree
in optical signal processing from the University of Essex, Colchester, U.K., in
2010. Between 2003 and 2009, he was a Senior Researcher with the Photonic Networks
Research Lab, University of Essex, U.K. After his Ph.D., he was appointed as a Senior Researcher with the Electro-optical communications group working on digital signal processing techniques and the development of space division multiplexed transmission (SDM) systems. 
He is currently an Associate Professor and leads the High-capacity optical transmission laboratory within the Institute for Photonic Integration (former COBRA), Department of Electrical Engineering, Eindhoven University of Technology (TU/e), The Netherlands. He was instrumental to the delivery of the first major SDM project in the European Union—MODEGAP project. Since 2014, he has been tenured at the ECO group, where he has since built up a world-class laboratory collaborating with several industrial and academic partners. His general research interests are in the areas of optical and digital signal processing, space division multiplexing techniques, and long-haul transmission techniques. In 2018, he was TPC chair for subcommittee 3 on digital signal handling. Between 2015 and 2017, he served on the TPC for the OSA conference on signal processing in photonic communications (SPPCom). In 2017 and 2018, he was the Program Chair and the Conference Chair at SPPCom, respectively. Dr. Okonkwo recently served as an associate editor for special edition of the IEEE Journal on Lightwave Technology. For the next 3 years, he has been retained to serve as technical programme subcommittee on Fiber-optic and waveguide devices and sensors (subcommittee D5) at Optical fiber communications conference OFC 2020-2022.
\end{IEEEbiographynophoto}

\begin{IEEEbiographynophoto}{Satoshi Shinada} (Member, IEEE)
received the B.S. degree from Science University of Tokyo in 1998 and the M.E. and Ph.D. degrees from Tokyo Institute of Technology in 2000 and 2002, respectively. In 2002 he joined the Precision and Intelligence Laboratory, Tokyo Institute of Technology as a JSPS Post-Doctoral Fellow. Since 2003, he has been with National Institute of Information and Communications Technology (NICT), Tokyo Japan. From 2015 to 2016, he was a Deputy Director of the Ministry of Internal Affairs and Communications, Japan. He has been engaged in the research and development on LiNbO3 optical modulators, optical switches, optical interfaces for single flux quantum circuit and optical packet switching (OPS) systems. Dr. Shinada received the IEEE/LEOS Student Award in 2002, and the 2015 Ichimura Prize in Science for Excellent Achievement from the New Technology Development Foundation. He is a member of IEEE, IEEE Photonics Society, the Japan Society of Applied Physics (JSAP) and the Institute of Electronics, Information and Communication Engineers of Japan (IEICE).
\end{IEEEbiographynophoto}

% \end{IEEEbiography}


%\newpage



%\vfill


\end{document}